\documentclass[aps,prc,reprint,noshowpacs,groupedaddress,onecolumn1]{revtex4}
\usepackage{amsmath}
\usepackage{amssymb}
\usepackage{amsfonts}
\usepackage{amscd}
\usepackage{bm}
\usepackage{graphicx}
\usepackage{color}
\def\beq{\begin{equation}}
\def\eeq{\end{equation}}
\begin{document}

\title{Electromagnetic current operators for phenomenological relativistic models}

\author{W.~N.~Polyzou}
\affiliation{Department of Physics and Astronomy, The University of
Iowa, Iowa City, IA 52242}



\begin{abstract}

  {\bf Background:} Phenomenological Poincar\'e invariant quantum
  mechanical models can provide an efficient description of the dynamics
  of strongly interacting particles that is consistent with
  spectral and scattering observables.  These models are
  representation dependent and in order to apply them to
  reactions with electromagnetic probes it is necessary to have a
  consistent electromagnetic current operator.

  {\bf Purpose:} The purpose of this work is to use local gauge invariance
  to construct consistent strong current operators.

  {\bf Method:} Current operators are constructed from a model
  Hamiltonian by replacing momentum operators in the Weyl
  representation by gauge covariant derivatives.

  {\bf Results:} The construction provides a systematic method to construct
  explicit expressions for current operators that are consistent with
  relativistic models of strong interaction dynamics.
  
\end{abstract}

\maketitle

\section{Introduction}


Electromagnetic probes are useful tools for studying
strong interaction dynamics.  This is because they can be accurately
treated to lowest non-trivial order in the exchanged photon field.  In this
approximation scattering matrix elements are linear in matrix elements
of the electromagnetic current of the strongly interacting system.
Calculations of electromagnetic observables require consistent models
of the strong interaction dynamics and the hadronic current operators.

In quantum field theories the current is determined from the dynamics
by requiring local gauge invariance.  This is achieved by replacing
derivatives by gauge covariant derivatives and extracting
the coefficient of the term that is linear in the vector potential.

Phenomenological models provide an efficient means to model
the structure and dynamics of systems of
strongly interacting particles at energy and momentum scales
that require a relativistic treatment
\cite{Terent76}
\cite{Bereste76}
\cite{Glockle:1986zz}
\cite{PhysRevC.86.055205}
\cite{Keister:1988zz}
\cite{Chung:1988my}
\cite{Chung:1989zz}
\cite{Cardarelli:1995dc}
\cite{Polyzou:1996fb}
\cite{Schul96}
\cite{Krutov:1997wu}
\cite{Lev00}
\cite{Wagenbrunn:2000es}
\cite{JuliaDiaz:2003gq}
\cite{Coester:2005cv}
\cite{Lin:2007kg}
\cite{Lin:2008sy}
\cite{Huang:2008jd}
\cite{Fuda:2009zz}
\cite{Witala:2009zzb}
\cite{Desplanques:2009kj}
\cite{PhysRevC.102.065209}.

Inverse scattering methods \cite{Koshmanenko}
\cite{baumgartl:1983}(p.245) imply that in any finite energy interval
it is in principle possible to formulate phenomenological models
involving the experimentally accessible degrees of freedom that give
the same S-matrix elements as a more fundamental theory.  Cluster
properties \cite{Sokolov:1977}\cite{Coester:1982vt} provide a means to
build models of complex systems from models of few-body systems
that can be constrained by experiment.  One challenge is that the
relation of phenomenological Poincar\'e invariant models to an
underlying quantum field theory is neither direct nor straightforward.

The dynamics of free particles in relativistic quantum field theories
or relativistic quantum models is the same.  Free particles of mass
$m$ and spin $s$ transform as mass $m$ spin $s$ irreducible
representations of the Poincar\'e group.  In relativistic quantum
models with interactions there are short-range unitary transformations
that preserve the scattering matrix without modifying the
representation of the free particles that are used to formulate the
scattering asymptotic condition \cite{Ekstein:1960}.  S-matrix
preserving changes of representation of the dynamics require a
corresponding change in the representation of the hadronic current.
This is easy to understand; changes of representation change the wave
functions and hence the probability distribution of the charge
carrying constituents.  A corresponding change in the current is
needed to keep the electromagnetic observables invariant.  For
short-range unitary transformations the one-body parts of the current
remain unchanged, but the many-body parts are representation
dependent.

In principle dynamical many-body contributions to the current cannot be
ignored because current covariance and current conservation
cannot be satisfied without dynamical contributions to the current.
While these two conditions lead to representation-dependent dynamical
constraints on the hadronic current operator, they do not uniquely determine
the hadronic current operator, so additional information that depends on the
representation of the dynamics is absent.

It is possible to satisfy current covariance and current conservation
by calculating an independent set of current matrix elements in the
impulse approximation and use these constraints to generate the
remaining matrix elements.  This procedure is sensitive to the choice
of independent matrix elements and can only be applied to matrix
elements.  It assumes, without justification, that the required
many-body parts of the current do not contribute to the chosen set of
independent current matrix elements.  It is not clear how to
consistently use this method to treat reactions with different initial
and final states.  These considerations limit the use of
electromagnetic probes in relativistic quantum models to reactions
were the many-body parts of the hadronic current operator can be
ignored.  What is needed is a current operator that is consistent with
the dynamics, can be used for different initial and final states and
satisfies current covariance and current conservation at the operator level.

The requirement of local gauge invariance implies global gauge
invariance that results in current conservation.  In addition it
provides a means to construct the interaction-dependent part of the
strong current by replacing derivatives in the Hamiltonian associated
with electromagnetically interacting particles by gauge covariant
derivatives.  In what follows this will be done using the Weyl
representation of the dynamics.  This leads to a systematic method for
constructing a hadronic current operator that is consistent with the
model dynamics.

In this work, since Hamiltonians involve degrees of freedom at a
fixed time, the gauge invariance of the Hamiltonian is with respect to
space-dependent gauge transformations at a common fixed time.
The result of this construction is a current operator
that is constructed directly from the interaction.  Because the result
is an operator it has the advantage that it can be used with different
initial and final states.  Since the Hamiltonian involves
quantities at a fixed time, the resulting current is a 3-vector current;
however since the current transforms like a four vector density,
the dynamical part of the charge density operator can be expressed in terms
of the commutator of the dynamical rotationless boost generator with the
three-vector current.

While the construction presented in this work is systematic, preserves
local gauge invariance and results in an operator that is consistent
with the dynamics, the resulting current operator is not unique.  One
issue is the absence of a unique relativistic position operator.
While the momentum operator is $-i \times$ a partial derivative, the
derivative can be computed holding different types spin constant.
The natural choice is to compute the derivative holding the canonical
spin constant, but this is one of an infinite number of possibilities.
In addition the construction assumes that the interacting particles
are point charges.  This is not the case for nucleons, where finite
size effects result in one-body currents which depend on invariant
form factors.  The required form factors could be computed by applying
these methods to dynamical models based on sub-nucleon degrees of
freedom.

The general structure of relativistic quantum mechanical models of
particles is discussed in the next section.  The nature of the
representation dependence of hadronic current operators is also
discussed.  The formulation of gauge invariance in phenomenological
two-body Hamiltonians is discussed in section three.  The Weyl
representation of many-body operators is discussed in section four.
It expresses operators using the irreducible algebra of one-body
canonical coordinate and momentum operators, with the operators
ordered so the momentum operators are on the right of the coordinate
operators.  Gauge invariant extensions can be constructed by replacing
the momentum operators in the Weyl representation by gauge covariant
derivatives.  Section five illustrates the construction of a
dynamically generated two-body current using the example of the
$\mathbf{L}\cdot\mathbf{S}$, $(\mathbf{L}\cdot\mathbf{S})^2$ and
$\mathbf{L}\cdot\mathbf{L}$ parts of the non-relativistic Argonne V18
interaction.  The general construction, using the Weyl representation
is discussed in section six.  Section seven considers the case of
current operators generated by translationally invariant non-local
two-body interactions.  The general construction of section six is
applied to a square root kinetic energy operator in section eight.
The square root has non-localities that can be treated in the Weyl
representation.  The resulting current operator is compared to the
convection current for a free non-relativistic particle.  Both have
the form of charge times ``average velocity''.  The method of section
six results in a vector current density.  The associated charge
density is constructed from the vector current density using the
dynamical rotationless boost generator in section nine.  This ensures
that resulting 4-current density transforms like a Lorentz 4-vector
with respect to the dynamical representation of the Poincar\'e
group. The result is an operator rather than a matrix element of an
operator.  The gauge invariant form of the time-dependent
Schr\"odinger equation is discussed in section eleven.  The treatment
of the spin dependence in the relativistic case is discussed in
section twelve.  The application to relativistic light-front models
is also discussed in this section.  A summary of results and
conclusions are given in the final section.  There is one appendix
with a detailed construction of the $(\mathbf{L}\cdot\mathbf{S})^2$
and $\mathbf{L}\cdot\mathbf{L}$ parts of the current in section five.

\section{Electromagnetic currents in phenomenological models} 

A relativistic quantum mechanical model is defined by a unitary ray
representation, $U(\Lambda,a)$, of the component of the Poincar\'e
group continuously connected to the identity \cite{Wigner:1939cj}.
(Here $\Lambda$ labels a Lorentz transformation and $a$ labels the
displacement of a spacetime translation.)  This ensures that quantum
probabilities, expectation values, and ensemble averages are
independent of inertial reference frame.  This is a weaker condition
than microscopic locality, which can only be realized in theories with
an infinite number of degrees of freedom.

The relativistic analog of diagonalizing the Hamiltonian is
decomposing $U(\Lambda,a)$ into a direct integral of irreducible
representations.  Once the mass and spin operators are diagonalized,
the transformation properties of the states are fixed by group
theoretic considerations.

Conserved covariant current operators $J^{\mu}(x)$ satisfy
current conservation 
\[
[P^{\mu},J_{\mu}(x)]=0
\]
and current covariance
\[
U(\Lambda,a) J^{\mu}(x) U^{\dagger}(\Lambda,a)=(\Lambda^{-1})^\mu{}_{\nu}
J^{\nu}(\Lambda x+a )
\]
where $P^{\mu}$ is the generator of space-time translations.  Both of these
constraints involve the dynamics, since the Hamiltonian, $P^0$, and 
$U(\Lambda,a)$ are interaction dependent.

Unitary transformations $V$ satisfying
\beq
\lim_{t \to \pm \infty} \Vert V\Pi e^{-iH_0 t} \vert \psi \rangle \Vert =0,
\label{eq:em3}
\eeq
where $H_0$ is the asymptotic Hamiltonian and $\Pi$ is a channel projector 
result in a new $S$-matrix equivalent representation of the dynamics and the associated current:
\[
U'(\Lambda,a) = V U(\Lambda,a) V^{\dagger}
\]
\[
J^{\mu \prime}(x) = V J^{\mu}(x)V^{\dagger}. 
\]
The condition (\ref{eq:em3}), with both time limits, is necessary and
sufficient for $U(\Lambda,a)$ and $U'(\Lambda,a)$ to have the same $S$
matrix {\it without} changing the treatment of free particles
\cite{Ekstein:1960}\cite{PhysRevC.82.014002}.

While this $V$, which is a small perturbation of the identity in the
sense (\ref{eq:em3}), does not change the one-body parts of the
current operator, it changes the representation of the many-body
parts of the current.  What this means for phenomenological models is that the
representation of the current depends on the representation of
the dynamics.

\section{Gauge invariance in phenomenological models} 

While the focus in this work is on the two-body part of the hadronic
current, for the purpose of discussing gauge invariance it is useful
to express the dynamics in terms of fields.  For systems of a fixed
numbers of particles the fields are taken as the particle creation or
annihilation parts of relativistically covariant free nucleon fields
with physical masses.  These are not local operators since they do not
include the antiparticle operators.  They can be normalized to satisfy
equal time canonical commutation or anti-commutation relations.

A many-body Hamiltonian with two-body interactions can be
expressed in terms of these fields as 
\[
H = \int {\phi}^{\dagger}_a(\mathbf{q}',0) k_{ab}(\mathbf{q}',\mathbf{q})\phi_b (\mathbf{q},0)  d\mathbf{q}' d\mathbf{q} +
\]
\beq
\int 
{\phi}^{\dagger}_{b}(\mathbf{q}_b,0){\phi}^{\dagger}_{a}(\mathbf{q}_a,0)
v_{ab:cd}(\mathbf{q}_a,\mathbf{q}_b;\mathbf{q}_c,\mathbf{q}_d)
\phi_{c}(\mathbf{q}_c,0) \phi_{d}(\mathbf{q}_d',0)
 d\mathbf{q}_a  d\mathbf{q}_b  d\mathbf{q}_c  d\mathbf{q}_d 
\label{eq:d1} 
\eeq
where the coefficients   $k_{ab}(\mathbf{q}',\mathbf{q})$ and 
$v_{ab:cd}(\mathbf{q}_a,\mathbf{q}_b;\mathbf{q}_c,\mathbf{q}_d)$ are
matrix elements of the kinetic energy and interaction in the
single-particle basis used in the field and creation operators.

The spin degrees of freedom have been suppressed in these expressions.
This Hamiltonian is not necessarily local (in the sense of local
potentials) and is not invariant under local gauge transformations. It
is constructed to model the strong interaction dynamics for some range
of energies.

The Hamiltonian commutes with itself so it is independent of
time. This means that all of the fields in the Hamiltonian can be
evaluated at a fixed common time.

Under a local gauge transformation all of the field
operators are multiplied by position dependent phases, $e^{i\chi(\mathbf{q},t)}$,
at a fixed common
time.  The Hamiltonian (\ref{eq:d1}) will be gauge invariant if the phases
intertwine with the kernels in (\ref{eq:d1}) in the following sense
\beq
k_{ab}(\mathbf{q}',\mathbf{q})e^{i\chi(\mathbf{q},t)} = e^{i\chi(\mathbf{q}',t)}
k_{ab}(\mathbf{q}',\mathbf{q})
\label{eq:d2}
\eeq 
\beq
v_{ab:cd}(\mathbf{q}_a,\mathbf{q}_b;\mathbf{q}_c,\mathbf{q}_d) e^{i\chi(\mathbf{q}_c,t)+ i\chi(\mathbf{q}_d,t)}=
e^{i\chi(\mathbf{q}_a,t)+ i\chi(\mathbf{q}_b,t)} v_{ab:cd}(\mathbf{q}_a,\mathbf{q}_b;\mathbf{q}_c,\mathbf{q}_d). 
\label{eq:d3}
\eeq 
Satisfying this condition requires a modification of the original
Hamiltonian.  The modified Hamiltonian should become the original
Hamiltonian in the limit of 0 electric charge.


Since the interactions are non-local it is useful to consider
a geometric treatment of gauge transformations following \cite{Peskin}.
Under a local gauge transformation all of the charged
fields acquire spacetime dependent phases
\beq
\phi_a (\mathbf{q}_i,t) \to 
\phi_a' (\mathbf{q}_i,t) =
e^{i\chi(\mathbf{q},t)_i}\phi_a (\mathbf{q}_i,t) =: U(\mathbf{q}_i,t)
\phi_a(\mathbf{q}_i,t)
\label{eq:gt1}
\eeq
\beq
\phi_a^{\dagger}  (\mathbf{q}_i,t) \to 
\phi_a^{\dagger \prime} (\mathbf{q}_i,t) =
 \phi_a^{\dagger} (\mathbf{q}_i,t)
U^{\dagger}(\mathbf{q}_i,t) .
\label{eq:gt2}
\eeq
The derivatives of the gauge transformed fields transform like
\beq
-i \partial_{i\mu} \phi (\mathbf{q}_i,t) \to -i \partial_{i\mu} \phi' (\mathbf{q}_i,t) =
e^{i \chi(\mathbf{q}_i,t)} ( -i\partial_{i\mu}
+\partial_{i\mu}\chi_i(\mathbf{q}_i,t) ) \phi (\mathbf{q}_i,t).
\label{eq:gt3}
\eeq
This means that the derivative of the gauge transformed field is not
the gauge transformation of the derivative field.

While the Hamiltonian involves degrees of freedom at a common time,
the interactions are generally non-local, which implicitly involves
spatial derivatives.  The Hamiltonian can be transformed to a gauge invariant
operator by replacing the spatial derivatives by gauge covariant derivatives.

For the non-local case it is useful to
introduce a Wilson line operator $W(\mathbf{q}',t';\mathbf{q},t)$
that intertwines the gauge transformations at two different spacetime points
\beq
W(\mathbf{q}',t',\mathbf{q},t)U(\mathbf{q},t) = U(\mathbf{q}',t')
W(\mathbf{q}',t';\mathbf{q},t). 
\label{eq:gt4}
\eeq
For $\mathbf{q}'= \mathbf{q} +\delta \mathbf{q}$ close to $\mathbf{q}$
and fixed $t$
(\ref{eq:gt4}) becomes
\beq
W(\mathbf{q} +\delta \mathbf{q},t;\mathbf{q},t)U(\mathbf{q},t) = U(\mathbf{q} +\delta \mathbf{q},t) W(\mathbf{q} +\delta \mathbf{q},t;\mathbf{q},t).
\label{eq:gt5}
\eeq
This operator can be used to construct a gauge covariant
derivative. To do this consider the difference 
\beq
\phi_a (\mathbf{q} + \delta \mathbf{q},t) -
W(\mathbf{q}+\delta \mathbf{q},t;\mathbf{q},t) \phi_a(\mathbf{q},t).
\label{eq:gt6}
\eeq
Both terms in this equation undergo the same gauge
transformation when $\phi_a (\mathbf{q} + \delta \mathbf{q},t)$ and $\phi_a (\mathbf{q},t)$ are gauge transformed.  Dividing by $\delta \mathbf{q}$ and taking the limit as $\delta \mathbf{q} \to 0$
gives a gauge covariant derivative:
\[
\lim_{\delta \mathbf{q} \to 0}
  \frac{\phi_a (\mathbf{q} + \delta \mathbf{q},t) - (W(\mathbf{q}+\delta
      \mathbf{q},t;
      \mathbf{q},t) - W(\mathbf{q},t;\mathbf{q},t)+
      W(\mathbf{q},t;\mathbf{q},t))\phi_a(\mathbf{q}),t)}{\delta \mathbf{q}}=
\]
\[
\lim_{\delta \mathbf{q} \to 0} 
[\frac{\phi_a (\mathbf{q} + \delta \mathbf{q},t)- \phi_a(\mathbf{q},t)}{ \delta \mathbf{q}}
  - \frac{(W(\mathbf{q}+\delta \mathbf{q},t;\mathbf{q},t) - W(\mathbf{q},t;
    \mathbf{q},t)}{\delta \mathbf{q}}]\phi_a(\mathbf{q},t) = 
\]
\beq
(\pmb{\partial} -\pmb{\partial} W(\mathbf{q},t;\mathbf{q}),t)
\phi_a(\mathbf{q},t)
\label{eq:gt7}
\eeq
where the derivative in the second argument acts on the first $\mathbf{q}$
in $W(\mathbf{q},t;\mathbf{q}',t)$.
The vector potential is, up to a multiplicative constant, the term
linear in expanding $W(\mathbf{q}+\delta \mathbf{q},t;\mathbf{q},t)$ in powers of $\delta \mathbf{q}$:
\[
W(\mathbf{q}+\delta \mathbf{q},t;\mathbf{q},t) = 1 + \partial_{\mathbf{q}'}W(\mathbf{q}',t;\mathbf{q},t)_{\mathbf{q}'=\mathbf{q}}\cdot \delta \mathbf{q} +\cdots =
\]
\beq
1 -i e\mathbf{A}(\mathbf{q},t) \cdot \delta \mathbf{q} +
o(\mathbf{A}(\mathbf{q},t)^2) .
\label{eq:gt8}
\eeq

This result in the familiar expression for the covariant derivative
\beq
\mathbf{D} \phi_a(\mathbf{q},t) := \pmb{\partial} \phi_a(\mathbf{q},t)
-i e\mathbf{A}(\mathbf{q},t) \phi_a(\mathbf{q},t). 
\label{eq:gt9}
\eeq

\section{ Interactions - the Weyl representation}

The starting point for a general discussion of the dynamics is to note that
for a
system of a finite number of degrees of freedom any linear operator
$O$ can be represented in the Weyl representation \cite{Simon1} as
\beq
\hat{O}= \int d\mathbf{a} d\mathbf{b} o(\mathbf{a},\mathbf{b})
e^{i \mathbf{a}\cdot \hat{\mathbf{q}}} e^{i \mathbf{b}\cdot \hat{\mathbf{p}}}
\label{eq:w1}
\eeq
where in this section $\hat{\mathbf{p}}= (\hat{p}_1 , \cdots , \hat{p}_N)$ and $\hat{\mathbf{q}}
= (\hat{q}_1 , \cdots , \hat{q}_N)$ are conjugate
coordinate and momentum operators:
\beq
[\hat{q}_i,\hat{p}_j] = i \delta_{ij} .
\label{eq:w2}
\eeq
In this expression the hats distinguish operators from variables.
This follows from the Stone-Von Neumann theorem
\cite{VonNeumann2}\cite{Stone1}\cite{VonNeumann1}
\cite{simon}
which demonstrated the irreducibility of the Weyl algebra for systems
with finite number of degrees of freedom.  See \cite{SchwingerQM} for
an elementary treatment based on limits of finite systems.  For the
purpose of considering local gauge transformations the coordinates and conjugate momenta should be
organized into three-vectors, but in this section is it more efficient to
use a notation that groups the coordinates and momenta into a set of $N$ canonically
conjugate pairs of operators.  In a relativistic theory the single-particle
``coordinates'' are functions of the single-particle Poincar\'e
generators.  They can be taken as the one-body Newton-Wigner position operators,
\cite{Newton:1949cq} which in the single-particle momentum-canonical spin
representation are $i$ times the partial derivative with respect to
the single-particle momenta holding the single-particle canonical spin
constant (recall relativistic spins undergo momentum-dependent Wigner
rotations)\cite{Keister:1991sb}.  These operators are canonically
conjugate to the momentum generators and commute with the canonical spin.

The complex coefficients $o(\mathbf{a},\mathbf{b})$ can be expressed in
terms of matrix elements of the operator $\hat{O}$.  It is simplest to
start with a mixed (coordinate-momentum) representation where, due to the ordering of the operators in (\ref{eq:w1}), the
operators are replaced by numbers
\[
\langle \mathbf{q} \vert \hat{O} \vert \mathbf{p} \rangle =
\int d\mathbf{a} d\mathbf{b} o(\mathbf{a},\mathbf{b})
e^{i \mathbf{a}\cdot {\mathbf{q}}}\langle \mathbf{q} \vert \mathbf{p} \rangle e^{i \mathbf{b}\cdot {\mathbf{p}}}
\]
\beq
 =
\frac{1}{(2 \pi)^{N/2}} \int d\mathbf{a} d\mathbf{b} o(\mathbf{a},\mathbf{b})
e^{i \mathbf{a}\cdot \mathbf{q}}  
e^{i\mathbf{q}\cdot \mathbf{p}}e^{i \mathbf{b}\cdot \mathbf{p}} .
\label{eq:w3}
\eeq
This can be re-expressed in a form where the kernel $o(\mathbf{a},\mathbf{b})$
can be computed by Fourier transform:
\beq
(2 \pi)^{N/2} \langle \mathbf{q} \vert \hat{O} \vert \mathbf{p} \rangle
e^{-i\mathbf{q}\cdot \mathbf{p}} =
\int d\mathbf{a} d\mathbf{b} o(\mathbf{a},\mathbf{b})
e^{i \mathbf{a}\cdot \mathbf{q}} e^{i \mathbf{b}\cdot \mathbf{p}}.
\label{eq:w4}
\eeq
To extract the kernel multiply both sides by 
\beq
e^{-i \mathbf{a}'\cdot \mathbf{q}} e^{-i \mathbf{b}'\cdot \mathbf{p}} d\mathbf{p}
d\mathbf{q} 
\label{eq:w5}
\eeq
and integrate over $\mathbf{p}$ and $\mathbf{q}$ to get
\beq
(2 \pi)^{2N} o(\mathbf{a}',\mathbf{b}') =
\int (2 \pi)^{N/2} \langle \mathbf{q} \vert \hat{O} \vert \mathbf{p} \rangle
e^{-i(\mathbf{q}\cdot \mathbf{p}  + \mathbf{a}'\cdot \mathbf{q}
+ \mathbf{b}'\cdot \mathbf{p})}
d\mathbf{p}
d\mathbf{q}.
\label{eq:w6}
\eeq
This expresses $o(\mathbf{a}',\mathbf{b}')$ in terms of the coordinate-momentum matrix
elements of $O$:
\beq
o(\mathbf{a},\mathbf{b}) =
\int (2 \pi)^{N/2-2N} \langle \mathbf{q} \vert \hat{O} \vert \mathbf{p} \rangle
e^{-i(\mathbf{q}\cdot \mathbf{p}  + \mathbf{a}\cdot \mathbf{q}
+ \mathbf{b}\cdot \mathbf{p})}
d\mathbf{p}
d\mathbf{q}.
\label{eq:w7}
\eeq
Replacing the matrix element in the mixed representation by one in the
momentum representation gives
\[
o(\mathbf{a},\mathbf{b}) =
\]
\[
\int (2 \pi)^{N/2-2N} \langle \mathbf{q} \vert \mathbf{p}' \rangle
d\mathbf{p}' \langle \mathbf{p}' \vert \hat{O} \vert \mathbf{p} \rangle
e^{-i(\mathbf{q}\cdot \mathbf{p}  + \mathbf{a}\cdot \mathbf{q}
+ \mathbf{b}\cdot \mathbf{p})}
d\mathbf{p}
d\mathbf{q} =
\]
\beq
\int (2 \pi)^{-2N}
\langle \mathbf{p}' \vert \hat{O} \vert \mathbf{p} \rangle
e^{i(\mathbf{q}\cdot \mathbf{p}'  - \mathbf{a}\cdot \mathbf{q}
- \mathbf{b}\cdot \mathbf{p} - \mathbf{q}\cdot \mathbf{p})}
d\mathbf{p} d\mathbf{p}'
d\mathbf{q} .
\label{eq:w8}
\eeq
Integrating over $\mathbf{q}$ gives a delta function
\beq
\int (2 \pi)^{-N}
\langle \mathbf{p}' \vert \hat{O} \vert \mathbf{p} \rangle
e^{-i \mathbf{b}\cdot \mathbf{p}}
\delta (\mathbf{p}'-\mathbf{a}-\mathbf{p})
d\mathbf{p} d\mathbf{p}' .
\label{eq:w9}
\eeq
Integrating over $\mathbf{p}'$ gives
\beq
\int (2 \pi)^{-N}
\langle \mathbf{p}+\mathbf{a} \vert \hat{O} \vert \mathbf{p} \rangle
e^{-i \mathbf{b}\cdot \mathbf{p}}
d\mathbf{p}.
\label{eq:w10}
\eeq
This can be put in a more symmetric form by defining
$
\mathbf{p} = \mathbf{p}' - \mathbf{a}/2
$
\[
o(\mathbf{a},\mathbf{b}) =
\]
\beq
\int (2 \pi)^{-N}
\langle \mathbf{p}'+\mathbf{a}/2 \vert \hat{O} \vert \mathbf{p}'- \mathbf{a}/2 \rangle
e^{-i \mathbf{b}\cdot (\mathbf{p}'- \mathbf{a}/2)}
d\mathbf{p}'  .
\label{eq:w11}
\eeq
Removing the primes gives an expression for the coefficient $o(\mathbf{a},\mathbf{b})$ in terms of momentum-space matrix elements of $\hat{O}$: 
\beq
o(\mathbf{a},\mathbf{b}) =
\int (2 \pi)^{-N}
\langle \mathbf{p}+\mathbf{a}/2 \vert \hat{O} \vert \mathbf{p}- \mathbf{a}/2 \rangle
e^{-i \mathbf{b}\cdot (\mathbf{p}- \mathbf{a}/2)}
d\mathbf{p}.
\label{eq:w12}
\eeq
These relations are for a system of $N$ degrees of freedom.
They relate momentum-space matrix elements of operators to
the expansion coefficients of the operators in the Weyl representation.

For applications of interest $\hat{O}$ is a translationally
invariant two-body interaction.  For two particles equation (\ref{eq:w12})
becomes
\[
o(\mathbf{a}_1, \mathbf{a}_2,\mathbf{b}_1, \mathbf{b}_2) =
\]
\[
\int (2 \pi)^{-6}
\langle \mathbf{p}_1+\mathbf{a}_1/2,\mathbf{p}_2+\mathbf{a}_2/2 \vert \hat{O} \vert
\mathbf{p}_1- \mathbf{a}_1/2,\mathbf{p}_2- \mathbf{a}_2/2 \rangle
\times
\]
\beq
e^{-i \mathbf{b}_1\cdot (\mathbf{p}_1- \mathbf{a}_1/2)
-i \mathbf{b}_2\cdot (\mathbf{p}_2- \mathbf{a}_2/2)}
d\mathbf{p}_1 d\mathbf{p}_2.
\label{eq:w13}
\eeq
Translational invariance implies the operator $\hat{O}$
commutes with the total momentum so the matrix element in (\ref{eq:w13})
can be expressed as
\[
\delta (\mathbf{p}_1+\frac{\mathbf{a}_1}{2}+\mathbf{p}_2+\frac{\mathbf{a}_2}{2} -
\mathbf{p}_1+ \frac{\mathbf{a}_1}{ 2}-\mathbf{p}_2+\frac{\mathbf{a}_2}{ 2})
\langle \mathbf{p}_1+{\mathbf{a}_1}{ 2},\mathbf{p}_2+\frac{\mathbf{a}_2}{2} \Vert \hat{O} \Vert
\mathbf{p}_1- \frac{\mathbf{a}_1}{2},\mathbf{p}_2- \frac{\mathbf{a}_2}{2} \rangle =
\]
\beq
\delta (\mathbf{a}_1 +\mathbf{a}_2)
\langle \mathbf{p}_1+\frac{\mathbf{a}_1}{2},\mathbf{p}_2-\frac{\mathbf{a}_1}{2} \Vert \hat{O} \Vert
\mathbf{p}_1- \frac{\mathbf{a}_1}{2},\mathbf{p}_2+ \frac{\mathbf{a}_1}{2} \rangle .
\label{eq:w14}
\eeq

Since momentum is conserved it is useful to make the variable change
$\mathbf{P}= \mathbf{p}_1 + \mathbf{p}_2$ and
$\mathbf{p} = \frac{1}{2}(\mathbf{p}_1-\mathbf{p}_2)$,
$d\mathbf{p}_1 d\mathbf{p}_2= d\mathbf{P} d\mathbf{p}$.  
In terms of these variables the previous equation is replaced by
\beq
\delta (\mathbf{a}_1 +\mathbf{a}_2)
\langle \frac{1}{2}\mathbf{P} + \mathbf{p} +\mathbf{a}_1/2,
\frac{1}{2}\mathbf{P} - \mathbf{p}-\mathbf{a}_1/2 \Vert \hat{O} \Vert
\frac{1}{2}\mathbf{P} + \mathbf{p} - \mathbf{a}_1/2,\frac{1}{2}\mathbf{P} - \mathbf{p}+ \mathbf{a}_1/2 \rangle .
\label{eq:w15}
\eeq
If $\hat{O}$ is a self-adjoint operator it follows that
the Weyl coefficients satisfy
\beq
o(\mathbf{a},\mathbf{b}) = o^*(-\mathbf{a},-\mathbf{b})e^{i \mathbf{a}\cdot \mathbf{b} }.
\label{eq:w16}
\eeq

\section{Example}

It is useful to consider an example that illustrates the strategy for
the general method.  For simplicity this method will be applied to
a non-relativistic model.  For local potentials the potential
is locally gauge invariant. 
For a realistic  interaction like the Argonne V18
interaction \cite{Wiringa:1994wb} the operators that involve the
orbital angular momentum like spin orbit terms, $\mathbf{S}\cdot
\mathbf{L}$, $(\mathbf{S}\cdot \mathbf{L})^2$ and $L^2$ will lead to
interaction dependent currents, since the orbital angular momentum
involves derivatives.

The Weyl representation is useful because the operators are ordered so
the momentum operators are to the right of the position operators.  For
the Argonne V18 interaction the ordering can be more simply realized
by using canonical commutation relations, which necessarily gives the
same result as the Weyl representation. 

The two-body current associated with a non-relativistic spin-orbit
interaction is derived first.  In this case the derivative term appears
linearly.  The spin-orbit contribution to the current is derived by
replacing the momentum in the orbital angular momentum operator by a
gauge covariant derivative and extracting the coefficient of the
vector potential.

In what follows it is useful to define the following conjugate pairs of
momenta and coordinates
\[
\mathbf{P}= \mathbf{p}_1+\mathbf{p}_2 \qquad
\mathbf{p}= {1 \over 2} (\mathbf{p}_1 - \mathbf{p}_2)
\]
\beq
\mathbf{q}= \mathbf{q}_1-\mathbf{q}_2 \qquad
\mathbf{Q}= {1 \over 2} (\mathbf{q}_1+\mathbf{q}_2).
\label{ex:1}
\eeq

Assume a spin-orbit interaction of the form
\beq
v_{so}(\vert \mathbf{q}\vert  ) \mathbf{L}\cdot \mathbf{S} =
v_{so}(\vert \mathbf{q}\vert  ) (\mathbf{q}\times \mathbf{p})\cdot \mathbf{S}
=
v_{so}(\vert \mathbf{q}\vert  ) (\mathbf{S}\times \mathbf{q})\cdot \mathbf{p} 
\label{ex:2}
\eeq
where $\mathbf{L}= \mathbf{q} \times \mathbf{p}$.
For this interaction $[\mathbf{L},v_{so}(\vert \mathbf{q} \vert)]=0$ so $\mathbf{p}$ can be placed on the right of
$v_{so}$ as in the Weyl representation.  Replacing the
single particle momentum operators by covariant derivatives gives
\beq
v_{so}(\vert \mathbf{q}\vert  ) (\mathbf{S}\times \mathbf{q})\cdot
(\mathbf{p}  - {e_1 \over 2} \mathbf{A}(\mathbf{q}_1) +  {e_2 \over 2} \mathbf{A}(\mathbf{q}_2)).
\label{ex:3}
\eeq

Note that this replacement assumes that the nucleons are structureless
point charges with charge $e_i$.  Real nucleons have a finite size
with non-trivial charge and magnetic moment distributions.
The electromagnetic structure of individual nucleons does not directly impact
the structure of the nucleon-nucleon interaction, but it contributes to the
nuclear currents.  This will be ignored in this section, since the focus is on
the role of local gauge invariance in constructing consistent
two-body currents. 

In this case it is not necessary to use the Weyl representation
because (\ref{ex:2}) can be expressed directly with the momentum
operators to the right for the coordinate operators.

The part of (\ref{ex:3})  that is linear in the vector potential can be expressed as
\beq
\int \mathbf{J}(\mathbf{x},0) \cdot \mathbf{A}(\mathbf{x},0) d\mathbf{x} 
\label{ex:4}
\eeq
where
\beq
\mathbf{J}(\mathbf{x},0) =
- {1 \over 2} v_{so}(\vert \mathbf{q}\vert  ) (\mathbf{S}\times \mathbf{q})
(e_1\delta (\mathbf{x} -\mathbf{q}_1) -
e_2 \delta (\mathbf{x} -\mathbf{q}_2) ).
\label{ex:5}
\eeq
In this expression it is understood that the $\mathbf{q}_i$ are operators.
They become numbers after taking mixed basis matrix elements.
Taking momentum-space matrix elements of (\ref{ex:5}) gives
\[
\langle \mathbf{p}'_1, \mathbf{p}'_2 \vert \mathbf{J}(\mathbf{x},0)
\vert \mathbf{p}_1, \mathbf{p}_2 \rangle =
\]
\[
- {1 \over 2}
\int
\langle \mathbf{p}'_1, \mathbf{p}'_2 \vert \mathbf{Q}', \mathbf{q}' \rangle 
v_{so}(\vert \mathbf{q}\vert  ) (\mathbf{S}\times \mathbf{q})
\delta (\mathbf{Q}'-\mathbf{Q}) \times
\] 
\[
\delta (\mathbf{q}'-\mathbf{q})
(e_1\delta (\mathbf{x} -\mathbf{q}_1) -
e_2\delta (\mathbf{x} -\mathbf{q}_2))
d \mathbf{Q}d \mathbf{Q}' d \mathbf{q} d \mathbf{q}'
\langle \mathbf{Q}, \mathbf{q} \vert
\mathbf{p}_1, \mathbf{p}_2 \rangle = 
\]
\[
- {1 \over 2(2 \pi)^6}
\int
e^{-i ( \mathbf{p}'_1+ \mathbf{p}'_2 - \mathbf{p}_1- \mathbf{p}_2)\cdot \mathbf{Q}}
e^{-{i} \mathbf{q}\cdot (\mathbf{p}'-\mathbf{p})} 
v_{so}(\vert \mathbf{q}\vert  ) (\mathbf{S}\times \mathbf{q}) \times
\]
\[
(e_1\delta (\mathbf{x} -\mathbf{Q}-{1\over 2}\mathbf{q}) -
e_2\delta (\mathbf{x} -\mathbf{Q}+{1\over 2}\mathbf{q}))
d \mathbf{Q}d  \mathbf{q} =
\]
\[
- {e_1 \over 2(2 \pi)^6}
\int
e^{-i ( \mathbf{p}'_1+ \mathbf{p}'_2 - \mathbf{p}_1- \mathbf{p}_2)\cdot(  \mathbf{x}-{1 \over 2}\mathbf{q})}
e^{-{i} \mathbf{q}\cdot (\mathbf{p}'-\mathbf{p})} 
v_{so}(\vert \mathbf{q}\vert  ) (\mathbf{S}\times \mathbf{q}) d\mathbf{q}
\]
\[
+ {e_2 \over 2(2 \pi)^6}
\int
e^{-i ( \mathbf{p}'_1+ \mathbf{p}'_2 - \mathbf{p}_1- \mathbf{p}_2)\cdot
( \mathbf{x}+{1 \over 2}\mathbf{q})}
e^{-{i} \mathbf{q}\cdot (\mathbf{p}'-\mathbf{p})} 
v_{so}(\vert \mathbf{q}\vert  ) (\mathbf{S}\times \mathbf{q}) d\mathbf{q} =
\]
\[
{1 \over (2 \pi)^6} e^{-i (\mathbf{P}'-\mathbf{P})\cdot \mathbf{x}}
\int
[(e_1-e_2)\cos \left ({\mathbf{q}\cdot (\mathbf{P}'-\mathbf{P})\over 2}\right )
  -i (e_1+e_2)\sin \left ({\mathbf{q}\cdot (\mathbf{P}'-\mathbf{P})\over 2}\right )]  \times
\]
\beq
e^{-{i} \mathbf{q}\cdot (\mathbf{p}'-\mathbf{p})} 
v_{so}(\vert \mathbf{q}\vert  ) (\mathbf{S}\times \mathbf{q}) d\mathbf{q}.
\label{ex:6}
\eeq
Setting $\mathbf{x}=0$ this becomes
\[
\langle \mathbf{p}'_1, \mathbf{p}'_2 \vert \mathbf{J}(\mathbf{0},0)
\vert \mathbf{p}_1, \mathbf{p}_2 \rangle =
\]
\[
{1 \over (2 \pi)^6}
 \int
[(e_1-e_2)\cos \left ({\mathbf{q}\cdot (\mathbf{P}'-\mathbf{P})\over 2}\right )
  -i (e_1+e_2)\sin \left ({\mathbf{q}\cdot (\mathbf{P}'-\mathbf{P})\over 2}\right )] \times
\]
\beq
e^{-{i} \mathbf{q}\cdot (\mathbf{p}'-\mathbf{p})} 
v_{so}(\vert \mathbf{q}\vert  ) (\mathbf{S}\times \mathbf{q}) d\mathbf{q}. 
\label{ex:7}
\eeq
This operator has a non-trivial dependence on the interaction
$
v_{so}(\vert \mathbf{q}\vert  ) (\mathbf{S}\times \mathbf{q}) $.

Next consider operators of the form
$(\mathbf{L}\cdot \mathbf{S})^2$ and $(\mathbf{L}\cdot \mathbf{L})$.
These operators are quadratic in the momentum.

In order to carry out the above analysis the first step is to order
the operators so the coordinate operators are on the left of the
momentum operators.  This can be done using the Weyl representation,
but for these interactions a direct approach using the canonical
commutation relations can be used to keep the vector potential to the
left of the momentum operators.  The end result must be the same.
The steps are similar to the steps used with the spin-orbit interaction.
The resulting contributions to the current are:
\[
\langle \mathbf{P}',\mathbf{p'} \vert \mathbf{J}(\mathbf{x},0)_{(\mathbf{L}\cdot \mathbf{S})^2}
\vert \mathbf{P},\mathbf{p} \rangle=
\]
\[
{1 \over (2 \pi)^6} e^{i (\mathbf{P}-\mathbf{P}')\cdot \mathbf{x}}
\int
[
(e_1+e_2)\sin \left ({\mathbf{q}\cdot (\mathbf{P}'-\mathbf{P})\over 2}\right )
+i(e_1-e_2)\cos \left ({\mathbf{q}\cdot (\mathbf{P}'-\mathbf{P})\over 2}\right )
]
\times
\]
\beq
e^{i \mathbf{q}\cdot
(\mathbf{p}-\mathbf{p}')}v_{(\mathbf{L}\cdot \mathbf{S})^2}(\vert \mathbf{q} \vert)
\left [-{1 \over 2}   
(\mathbf{S} \times \mathbf{q})\cdot (\mathbf{p}+\mathbf{p}')(\mathbf{S} \times \mathbf{q}))
-{i \over 2}  (\mathbf{S}^2 \mathbf{q} - (\mathbf{S}\cdot \mathbf{q})\mathbf{S}) 
\right ] d\mathbf{q}
\label{ex:8}
\eeq
and
\[
\langle \mathbf{P}',\mathbf{p}' \vert 
\mathbf{J}(\mathbf{x},0)_{\mathbf{L}\cdot \mathbf{L}}
\vert \mathbf{P},\mathbf{p} \rangle =
\]
\[
\int d\mathbf{q} 
v_{\mathbf{L}\cdot \mathbf{L}}(\vert \mathbf{q} \vert)
{1 \over (2\pi)^6} e^{i \mathbf{x}\cdot (\mathbf{P}-\mathbf{P}') + i \mathbf{q}\cdot
(\mathbf{p}-\mathbf{p}')} \times
\]
\[
[ (e_1-e_2) \cos \left ({\mathbf{q}\cdot (\mathbf{P}-\mathbf{P}'\over 2} \right) +i (e_1+e_2) 
\sin \left ({\mathbf{q}\cdot (\mathbf{P}-\mathbf{P}'\over 2} \right )]
\times
\]
\beq
{1 \over 2} \mathbf{q} \times (\mathbf{q} \times (\mathbf{p}+\mathbf{p}'))
\label{ex:9}
\eeq

The derivations are given in the appendix.

\section{Gauge covariant non-local operators}

A general two-body Hamiltonian of the form (\ref{eq:d1}) will not be
locally gauge invariant in the sense (\ref{eq:d2}-\ref{eq:d3}). A
gauge invariant extension of (\ref{eq:d1}) can be constructed by
replacing the momentum operators in the Weyl representation by gauge
covariant derivatives:
\beq
\langle \mathbf{q}',t \vert e^{i \hat{\mathbf{p}} \cdot \mathbf{a} }\vert \mathbf{q}'',t \rangle
\to
\langle \mathbf{q}',t\vert e^{i (\hat{\mathbf{p}}-eA(\hat{\mathbf{q}},t)) \cdot \mathbf{a} }\vert \mathbf{q}'' ,t\rangle .
\label{eq:nl1}
\eeq

To show this has the desired property use the Trotter product
representation \cite{Simon1} of the exponent 
of the sum of non-commuting operators:
\beq
\lim_{N \to \infty} \langle \mathbf{q}' \vert
[ e^{i (\hat{\mathbf{p}}-eA(\hat{\mathbf{q}})) \cdot {\mathbf{a}\over N} }]^N\vert \mathbf{q}'' \rangle .
\label{eq:nl2}
\eeq
The important thing about this expression is that in the $N\to \infty$
limit only the
first order terms in the expansion of the exponential contribute.
It follows that (\ref{eq:nl2}) is equal to
\[
=\lim_{N \to \infty} \langle \mathbf{q}' ,t\vert
[1+i (\hat{\mathbf{p}}-eA(\hat{\mathbf{q}},t)) \cdot {\mathbf{a}\over N} ]^N
\vert \mathbf{q}'',t \rangle =
\]  
\[
\lim_{N \to \infty} \langle \mathbf{q}',t \vert
[(1+i \hat{\mathbf{p}} {\mathbf{a}\over N})
( 1-ieA(\hat{\mathbf{q}},t)) \cdot {\mathbf{a}\over N})
]^N\vert \mathbf{q}'' \rangle =
\]
\beq
\lim_{N \to \infty} \langle \mathbf{q}',t \vert
[e^{i \hat{\mathbf{p}} {\mathbf{a}\over N}}
e^{-ieA(\hat{\mathbf{q}},t) \cdot {\mathbf{a}\over N}}
]^N\vert \mathbf{q}'',t \rangle
\label{eq:nl3}
\eeq
where 
\beq
e^{i \hat{\mathbf{p}} {\mathbf{a}\over N}} = \hat{T}({\mathbf{a} \over N})
\label{eq:nl4}
\eeq
is the operator that translates the coordinates by ${\mathbf{a}\over N}$ and
\beq
e^{-ieA(\hat{\mathbf{q}},t) \cdot {\mathbf{a}\over N}} = \hat{W} (\mathbf{q}+{\mathbf{a}\over N},\mathbf{q})
\label{eq:nl5}
\eeq
is the operator that transforms the phase at $\mathbf{q}$ to the
phase at $\mathbf{q}+\mathbf{a}/N$ for small $\mathbf{a}/N$ (see \ref{eq:gt8}))..
With these identifications, inserting complete sets of intermediate states
expression (\ref{eq:nl3}) becomes
\[
\lim_{N \to \infty} \langle \mathbf{q}',t \vert
[e^{i \hat{\mathbf{p}} {\mathbf{a}\over N}}
e^{-ieA(\hat{\mathbf{q}},t) \cdot {\mathbf{a}\over N}}
]^N\vert \mathbf{q}'',t \rangle =
\]
\[
\int
\langle \mathbf{q}+\mathbf{a},t \vert
\hat{T}({\mathbf{a}\over N})
\hat{W}(\mathbf{q}+\mathbf{a},t, \mathbf{q}+ \mathbf{a}({N-1\over N}),t)
\vert \mathbf{q}_{N-1} \rangle d\mathbf{q}_{N-1} \times
\]
\[\langle \mathbf{q}_{N-1},t \vert 
\hat{T}({\mathbf{a}\over N})
\hat{W}( \mathbf{q}+\mathbf{a}{N-1 \over N},t; \mathbf{q}+\mathbf{a}{N-2 \over N})t
\vert \mathbf{q}_{N-2} ,t\rangle d\mathbf{q}_{N-2} 
\langle d\mathbf{q}_{N-2}
\]
\[
\cdots
\vert \mathbf{q}_2 \rangle d\mathbf{q}_2 \langle \mathbf{q}_2 \vert 
\hat{T}({\mathbf{a}\over N})
\hat{W}(\mathbf{q}+2{\mathbf{a}\over N},t;\mathbf{q}+{\mathbf{a}\over N},t)
\vert \mathbf{q}_1 \rangle \times
\]
\beq
d\mathbf{q}_1 \langle \mathbf{q}_1 \vert 
\hat{T}({\mathbf{a} \over N})
\hat{W}(\mathbf{q}+{\mathbf{a}\over N},t;\mathbf{q},t)\vert \mathbf{q}'',t \rangle . 
\label{eq:nl6}
\eeq
After a local gauge transform on the initial coordinate, 
each step translates the coordinate by $\mathbf{a}/N$ and transports the
phase to the phase associated with the translated coordinate.
This implies that in the limit
$N\to \infty$  (\ref{eq:nl6}) has the
property \cite{Partovi}
\[
\langle \mathbf{q}'+\mathbf{a},t \vert e^{i (\hat{\mathbf{p}}-eA(\hat{\mathbf{q}},t)) \cdot \mathbf{a} }\vert \mathbf{q}' \rangle e^{i\chi(\mathbf{q}',t)} =
\]
\beq
e^{i\chi(\mathbf{q}'+\mathbf{a},t)}
\langle \mathbf{q}'+\mathbf{a} \vert e^{i (\hat{\mathbf{p}}-eA(\hat{\mathbf{q}})) \cdot \mathbf{a} }\vert \mathbf{q}' \rangle .
\label{eq:n17}
\eeq
The operator $e^{i (\hat{\mathbf{p}}-eA(\hat{\mathbf{q}})) \cdot \mathbf{a} } = e^{i \mathbf{D}\cdot\mathbf{a}}$ in
(\ref{eq:n17}) is
the exponential of the gauge covariant operator $\mathbf{D}\cdot\mathbf{a}$.

This shows that
\beq
\hat{O} = \int d\mathbf{a} d \mathbf{b} o(\mathbf{a},\mathbf{b})
e^{i \mathbf{a}\cdot \hat{\mathbf{q}}}
e^{i (\hat{\mathbf{p}}-eA(\hat{\mathbf{q}})) \cdot \mathbf{b} }
\label{eq:nl8}
\eeq
is a gauge covariant operator (in the relativistic case $\hat{\mathbf{q}}$ represents a
single-particle Newton-Wigner position operator).

Note that if the original $\hat{O}$ is self-adjoint then so is the
corresponding gauge covariant operator.  This follows because
\[
[e^{i \mathbf{a}\cdot \hat{\mathbf{q}}}
e^{i (\hat{\mathbf{p}}-eA(\hat{\mathbf{q}})) \cdot \mathbf{b} }]^{\dagger} =
e^{-i (\hat{\mathbf{p}}-eA(\hat{\mathbf{q}})) \cdot \mathbf{b} }
e^{-i \mathbf{a}\cdot \hat{\mathbf{q}}}=
\]
\beq
e^{-i \mathbf{a}\cdot \hat{\mathbf{q}}}
e^{-i (\hat{\mathbf{p}}-\mathbf{a}-eA(\hat{\mathbf{q}})) \cdot \mathbf{b}} =
e^{-i \mathbf{a}\cdot \hat{\mathbf{q}}}
e^{-i (\hat{\mathbf{p}}-eA(\hat{\mathbf{q}}))\cdot \mathbf{b}}
e^{i \mathbf{a}\cdot \mathbf{b}}
\label{eq:nl9}
\eeq
which shows that the gauge covariant operator is self adjoint
if the coefficients $o(\mathbf{a},\mathbf{b})$ satisfy 
condition (\ref{eq:w16}).

\section{Two-body vector currents}

The interesting quantity in the one-photon exchange approximation is
the hadronic current which is the coefficient of the part of this
gauge-invariant Hamiltonian that is linear in the gauge field.  The
structure of the interaction-dependent two-body contribution to this
current is discussed in this section.

A two-body interaction in the Weyl representation has the form
\beq
\hat{v}= \int d\mathbf{a} d\mathbf{b} v(\mathbf{a}_1,\mathbf{a}_2,\mathbf{b}_1,\mathbf{b}_2)
e^{i( \mathbf{a}_1\cdot \hat{\mathbf{q}}_1+  \mathbf{a}_2\cdot \hat{\mathbf{q}}_2)
}
e^{i( \mathbf{b}_1\cdot \hat{\mathbf{p}}_1+  \mathbf{b}_2\cdot \hat{\mathbf{p}}_2)
}.
\label{eq:tbc1}
\eeq
Replacing the momentum operators by covariant derivatives in the Weyl
representation results in the following
gauge covariant kernel for the two-body interaction:
\[
\hat{v}_g= \int   d\mathbf{a}_1 d\mathbf{a}_2 d\mathbf{b}_1 d\mathbf{b}_2
v(\mathbf{a}_1,\mathbf{a}_2,\mathbf{b}_1,\mathbf{b}_2)
e^{i( \mathbf{a}_1\cdot \hat{\mathbf{q}}_1+  \mathbf{a}_2\cdot \hat{\mathbf{q}}_2) }
\times
\]
\beq
e^{i(( \mathbf{b}_1\cdot (\hat{\mathbf{p}}_1
  - {e_1 } \mathbf{A}(\hat{\mathbf{q}}_1,t)) +
   (\mathbf{b}_2\cdot (\hat{\mathbf{p}}_2
  - {e_2} \mathbf{A}(\hat{\mathbf{q}}_2,t)))}.
\label{eq:tbc2}
\eeq

The two-body part of the current operator is the coefficient
of the part of $\hat{v}_g$ that is linear in the charges $e_i$.  Since the
momenta and vector potential do not commute, in order to find the term
linear in the $e_i$ use for $B$ and $A$ non-commuting operators
\beq
{d\over dx}e^{\hat{B}+\hat{A}x}_{\vert x=0} = \int_0^1 d\lambda
e^{\lambda \hat{B}}\hat{A} e^{(1-\lambda)\hat{B}} .
\label{eq:tbc3}
\eeq
This formula follows using
the Trotter product formula 
\beq
e^{\hat{A}+\hat{B}} \vert \Psi \rangle = \lim_{N\to \infty} 
(e^{\hat{A}/N}e^{\hat{B}/N})^N \vert \Psi \rangle
\eeq
and the chain rule \cite{Georgi}.
The part of the interaction that is linear in the vector potential is 
\[
\langle \mathbf{q}_2, \mathbf{q}_1 \vert \sum_i e_i{\partial \hat{v}_g \over \partial e_i} \vert
\mathbf{p}_1,\mathbf{p}_2 \rangle_{e_i=0} =
\]  
\[
\int d\mathbf{a}_1d\mathbf{a}_2d\mathbf{b}_1d\mathbf{b}_2 \int_0^1d\lambda
v(\mathbf{a}_1,\mathbf{a}_2;\mathbf{b}_1,\mathbf{b}_2)
e^{i( \mathbf{a}_1\cdot\hat{\mathbf{q}}_1+
  \mathbf{a}_2\cdot\hat{\mathbf{q}}_2) }
\]
\beq
e^{i \lambda (  \mathbf{b}_1 \cdot \hat{\mathbf{p}}_1 +
\mathbf{b}_2\cdot\hat{\mathbf{p}}_2) }
( - ie_1 \mathbf{A}(\hat{\mathbf{q}}_1,t)\cdot \mathbf{b}_1
-i e_2\mathbf{A}(\hat{\mathbf{q}}_2,t)\cdot \mathbf{b}_2)
e^{i (1-\lambda) (\mathbf{b}_1 \cdot \hat{\mathbf{p}}_1+
\mathbf{b}_2 \cdot \hat{\mathbf{p}}_2)} =
\label{eq:tbc4}
\eeq

\[
\int d\mathbf{a}_1d\mathbf{a}_2d\mathbf{b}_1d\mathbf{b}_2
\int_0^1d\lambda
v(\mathbf{a}_1,\mathbf{a}_2;\mathbf{b}_1,\mathbf{b}_2)
e^{i( \mathbf{a}_1\cdot\hat{\mathbf{q}}_1+
  \mathbf{a}_2\cdot\hat{\mathbf{q}}_2) }
\]
\beq
( -ie_1 \mathbf{A}(\hat{\mathbf{q}}_1+\lambda \mathbf{b}_1 ,t)\cdot \mathbf{b}_1
-ie_2 \mathbf{A}(\hat{\mathbf{q}}_2 +\lambda \mathbf{b}_2 ,t) \cdot \mathbf{b}_2
e^{i(  \mathbf{b}_1 \cdot{\mathbf{p}}_1 +
\mathbf{b}_2\cdot {\mathbf{p}}_2)} +o(e^2) .
\label{eq:tbc6}
\eeq

In this expression 
the  $\hat{\mathbf{q}}_i$ are operators and all of the
$\hat{\mathbf{q}}_i$'s  are to the left of the $\hat{\mathbf{p}}_i$'s.
By taking mixed
matrix elements the operators become numbers
\[
\sum_i e_i \langle \mathbf{q}_2, \mathbf{q}_1 \vert {\partial \hat{v}_g \over \partial e_i} \vert
\mathbf{p}_1,\mathbf{p}_2 \rangle =
\]
\[
\int d\mathbf{a}_1d\mathbf{a}_2d\mathbf{b}_1d\mathbf{b}_2
\int_0^1d\lambda
v(\mathbf{a}_1,\mathbf{a}_2;\mathbf{b}_1,\mathbf{b}_2)
e^{i( \mathbf{a}_1\cdot{\mathbf{q}}_1+
\mathbf{a}_2\cdot{\mathbf{q}}_2) } \times
\]
\[
( -ie_1 \mathbf{A}({\mathbf{q}}_1+\lambda\mathbf{b}_1,t)\cdot \mathbf{b}_1
-ie_2 \mathbf{A}({\mathbf{q}}_2+\lambda\mathbf{b}_2,t)\cdot \mathbf{b}_2
) \times
\]
\beq
\langle \mathbf{q}_1 ,\mathbf{q}_2 \vert \mathbf {p}_1, \mathbf{p}_2 \rangle
e^{i(  \mathbf{b}_1 \cdot{\mathbf{p}}_1 +
\mathbf{b}_2\cdot {\mathbf{p}}_2)} +o(e^2) .
\label{eq:tbc7}
\eeq
To factor out the vector potential insert delta functions 
\[
\int d\mathbf{x} (-i \mathbf{A}({\mathbf{x}}))\cdot
\int d\mathbf{a}_1d\mathbf{a}_2d\mathbf{b}_1d\mathbf{b}_2 
\int_0^1d\lambda  
v(\mathbf{a}_1,\mathbf{a}_2;\mathbf{b}_1,\mathbf{b}_2)
\times
\]
\[
e^{i( \mathbf{a}_1\cdot{\mathbf{q}}_1+
  \mathbf{a}_2\cdot{\mathbf{q}}_2) }
(e_1 \delta (\mathbf{x}-\mathbf{q}_1-\lambda \mathbf{b}_1)\mathbf{b}_1 +
e_2\delta (\mathbf{x}-\mathbf{q}_2-\lambda \mathbf{b}_2)\mathbf{b}_2)
\times
\]
\beq
\langle \mathbf{q}_1 ,\mathbf{q}_2 \vert \mathbf {p}_1, \mathbf{p}_2 \rangle
e^{i(  \mathbf{b}_1 \cdot {\mathbf{p}}_1 +
  \mathbf{b}_2\cdot{\mathbf{p}}_2)} +o(e^2)  =
\label{eq:tbc8}
\eeq
\[
{1 \over (2\pi)^3} 
\int d\mathbf{x} (-i \mathbf{A}({\mathbf{x}}))\cdot
\int d\mathbf{a}_1d\mathbf{a}_2d\mathbf{b}_1d\mathbf{b}_2 
\int_0^1d\lambda  
v(\mathbf{a}_1,\mathbf{a}_2;\mathbf{b}_1,\mathbf{b}_2)
\times
\]
\[
e^{i( \mathbf{a}_1\cdot{\mathbf{q}}_1+
  \mathbf{a}_2\cdot{\mathbf{q}}_2) } \times
(e_1 \delta (\mathbf{x}-\mathbf{q}_1-\lambda \mathbf{b}_1)\mathbf{b}_1 +
e_2
\delta (\mathbf{x}-\mathbf{q}_2-\lambda \mathbf{b}_2)\mathbf{b}_2 )\times
\]
\beq
e^{i(  \mathbf{q}_1 \cdot {\mathbf{p}}_1 +
  \mathbf{q}_2\cdot{\mathbf{p}}_2)}
e^{i(  \mathbf{b}_1 \cdot {\mathbf{p}}_1 +
  \mathbf{b}_2\cdot {\mathbf{p}}_2)}  +o(e^2) .
\label{eq:tbc9}
\eeq
From this expression matrix elements of the current density in the mixed representation can be read off:  
\[
\langle \mathbf{q}_1,\mathbf{q}_2 \vert \mathbf{J}(\mathbf{x},0)
\vert \mathbf{p}_1 ,\mathbf{p}_2 \rangle:=
\]
\[
-{i \over (2\pi)^3} 
\int d\mathbf{a}_1d\mathbf{a}_2d\mathbf{b}_1d\mathbf{b}_2 
\int_0^1d\lambda  
v(\mathbf{a}_1,\mathbf{a}_2;\mathbf{b}_1,\mathbf{b}_2)
e^{i( \mathbf{a}_1\cdot{\mathbf{q}}_1+
  \mathbf{a}_2\cdot{\mathbf{q}}_2) } \times
\]
\beq
(e_1 \delta (\mathbf{x}-\mathbf{q}_1-\lambda \mathbf{b}_1)\mathbf{b}_1 +
e_2 \delta (\mathbf{x}-\mathbf{q}_2-\lambda \mathbf{b}_2)\mathbf{b}_2 )
e^{i(  \mathbf{q}_1 \cdot {\mathbf{p}}_1 +
  \mathbf{q}_2\cdot{\mathbf{p}}_2)}
e^{i(  \mathbf{b}_1 \cdot {\mathbf{p}}_1 +
  \mathbf{b}_2\cdot {\mathbf{p}}_2)}.
\label{eq:tbc9a}
\eeq
This can be Fourier transformed to give a momentum space kernel
\[ 
\langle \mathbf{p}_2',\mathbf{p}'_1 \vert \mathbf{J}(\mathbf{x},0)
\vert \mathbf{p}_1 ,\mathbf{p}_2 \rangle :=
\]
\[
-{i \over (2\pi)^6} 
\int d\mathbf{a}_1d\mathbf{a}_2d\mathbf{b}_1d\mathbf{b}_2d\mathbf{q}_1
d\mathbf{q}_2 \int_0^1d\lambda
v(\mathbf{a}_1,\mathbf{a}_2;\mathbf{b}_1,\mathbf{b}_2)
e^{i( \mathbf{a}_1\cdot{\mathbf{q}}_1+
\mathbf{a}_2\cdot{\mathbf{q}}_2) } \times
\]
\[
( e_1\delta (\mathbf{x}-\mathbf{q}_1-\lambda \mathbf{b}_1)\mathbf{b}_1 +
e_2 \delta (\mathbf{x}-\mathbf{q}_2-\lambda \mathbf{b}_2)\mathbf{b}_2 )
\times
\]
\beq
e^{i(  \mathbf{q}_1 \cdot {\mathbf{p}}_1 +
  \mathbf{q}_2\cdot{\mathbf{p}}_2)}
e^{i(  \mathbf{b}_1 \cdot \mathbf{p}_1 +
  \mathbf{b}_2\cdot \mathbf{p}_2)}
e^{-i(  \mathbf{q}_1 \cdot \mathbf{p}'_1 +
  \mathbf{q}_2\cdot \mathbf{p}'_2)}.
\label{eq:tbc10}
\eeq
One of the two $\mathbf{q}$ integrals can be performed using the delta functions
giving
\[
\langle \mathbf{p}_2',\mathbf{p}'_1 \vert \mathbf{J}(\mathbf{x},0)
\vert \mathbf{p}_1 ,\mathbf{p}_2 \rangle 
=-{i e_1 \over (2\pi)^6} 
\int d\mathbf{a}_1d\mathbf{a}_2d\mathbf{b}_1d\mathbf{b}_2
d\mathbf{q}_2\int_0^1d\lambda \times
\]
\[
v(\mathbf{a}_2,\mathbf{a}_1;\mathbf{b}_1,\mathbf{b}_2)
e^{i( \mathbf{a}_1\cdot (\mathbf{x}-\lambda \mathbf{b}_1)+
\mathbf{a}_2\cdot{\mathbf{q}}_2) } \times
\]
\[
\mathbf{b}_1
e^{i(  (\mathbf{x}-\lambda \mathbf{b}_1) \cdot {\mathbf{p}}_1 +
  \mathbf{q}_2\cdot{\mathbf{p}}_2)}
e^{i(  \mathbf{b}_1 \cdot \mathbf{p}_1 +
  \mathbf{b}_2\cdot \mathbf{p}_2)}
e^{-i((\mathbf{x}-\lambda \mathbf{b}_1) \cdot \mathbf{p}'_1 +
  \mathbf{q}_2\cdot \mathbf{p}'_2)} 
\]
\[
-{ie_2 \over (2\pi)^6} 
\int d\mathbf{a}_1d\mathbf{a}_2d\mathbf{b}_1d\mathbf{b}_2d\mathbf{q}_1
\int_0^1d\lambda \times
\]
\[
v(\mathbf{a}_1,\mathbf{a}_2;\mathbf{b}_1,\mathbf{b}_2)
e^{i( \mathbf{a}_1\cdot{\mathbf{q}}_1+
\mathbf{a}_2\cdot (\mathbf{x}-\lambda \mathbf{b}_2) } \times
\]
\beq
\mathbf{b}_2
e^{i(  \mathbf{q}_1 \cdot {\mathbf{p}}_1 +
  (\mathbf{x}-\lambda \mathbf{b}_2)\cdot{\mathbf{p}}_2)}
e^{i(  \mathbf{b}_1 \cdot \mathbf{p}_1 +
  \mathbf{b}_2\cdot \mathbf{p}_2)}
e^{-i(  \mathbf{q}_1 \cdot \mathbf{p}'_1 +
  (\mathbf{x}-\lambda \mathbf{b}_2)\cdot \mathbf{p}'_2)} .
\label{eq:tbc12}
\eeq
Integrating over the remaining $\mathbf{q}_2$ (resp $\mathbf{q}_1$) gives $(2\pi)^3 \times $
delta functions: 
\[
\langle \mathbf{p}_2',\mathbf{p}'_1 \vert \mathbf{J}(\mathbf{x},0)
\vert \mathbf{p}_1 ,\mathbf{p}_2 \rangle 
=-{i e_1\over (2\pi)^3} 
\int d\mathbf{a}_1d\mathbf{a}_2d\mathbf{b}_1d\mathbf{b}_2
\int_0^1d\lambda \times
\]
\[
v(\mathbf{a}_1,\mathbf{a}_2;\mathbf{b}_1,\mathbf{b}_2)
\delta (\mathbf{a}_2 +\mathbf{p}_2 - \mathbf{p}_2')
e^{i \mathbf{a}_1\cdot (\mathbf{x}-\lambda \mathbf{b}_1)} 
\mathbf{b}_1
e^{i (\mathbf{x}-\lambda \mathbf{b}_1) \cdot {\mathbf{p}}_1} 
e^{i(  \mathbf{b}_1 \cdot \mathbf{p}_1 +
  \mathbf{b}_2\cdot \mathbf{p}_2)}
e^{-i(\mathbf{x}-\lambda \mathbf{b}_1) \cdot \mathbf{p}'_1} +
\]
\[
-{ie_2 \over (2\pi)^3} 
\int d\mathbf{a}_1d\mathbf{a}_2d\mathbf{b}_1d\mathbf{b}_2
\int_0^1d\lambda \times
\]
\beq
v(\mathbf{a}_1,\mathbf{a}_2;\mathbf{b}_1,\mathbf{b}_2)
\delta (\mathbf{a}_1 +\mathbf{p}_1 - \mathbf{p}_1')
e^{i
\mathbf{a}_2\cdot (\mathbf{x}-\lambda \mathbf{b}_2) } 
\mathbf{b}_2
e^{i(
  \mathbf{x}-\lambda \mathbf{b}_2)\cdot{\mathbf{p}}_2}
e^{i(  \mathbf{b}_1 \cdot \mathbf{p}_1 +
  \mathbf{b}_2\cdot \mathbf{p}_2)}
e^{-i(
  \mathbf{x}-\lambda \mathbf{b}_2)\cdot \mathbf{p}'_2}  .
\label{eq:tbc13}
\eeq
Integrating $\mathbf{a}_2= \mathbf{p}_2'-\mathbf{p}_2$ resp $\mathbf{a}_1=\mathbf{p}_1'-\mathbf{p}_1$ using the delta functions gives:
\[
\langle \mathbf{p}_2',\mathbf{p}'_1 \vert \mathbf{J}(\mathbf{x},0)
\vert \mathbf{p}_1 ,\mathbf{p}_2 \rangle =
-{ie_1 \over (2\pi)^3} \int d\mathbf{a}_1d\mathbf{b}_1d\mathbf{b}_2
 \int_0^1d\lambda \times
\]
\[
v(\mathbf{a}_1,(\mathbf{p}_2'-\mathbf{p}_2);\mathbf{b}_1,\mathbf{b}_2)
\mathbf{b}_1 e^{i \mathbf{x}\cdot (\mathbf{a}_1+ \mathbf{p}_1
  -\mathbf{p}_1')} e^{i \mathbf{b}_2 \cdot \mathbf{p}_2 } e^{i
  \mathbf{b}_1 \cdot ( -\lambda \mathbf{a}_1-\lambda
  \mathbf{p}_1+\lambda \mathbf{p}_1' +\mathbf{p}_1)}
\]
\[
-{ie_2 \over (2\pi)^3} \int
d\mathbf{a}_2d\mathbf{b}_1d\mathbf{b}_2 \int_0^1d\lambda \times
\]
\beq
v((\mathbf{p}_1'-\mathbf{p}_1),\mathbf{a}_2;\mathbf{b}_1,\mathbf{b}_2)
\mathbf{b}_2 e^{i \mathbf{x}\cdot (\mathbf{a}_2+ \mathbf{p}_2
  -\mathbf{p}_2')} e^{i \mathbf{b}_1 \cdot \mathbf{p}_1 } e^{i
  \mathbf{b}_2 \cdot ( -\lambda \mathbf{a}_2-\lambda
  \mathbf{p}_2+\lambda \mathbf{p}_2' +\mathbf{p}_2)}.
\label{eq:tbc14}
\eeq
The next step is to replace the kernel of the interaction in the Weyl
representation by its expression in terms of
potential matrix elements using
\[
v(\mathbf{a}_1,\mathbf{a}_2;\mathbf{b}_1,\mathbf{b}_2) =
\]
\[
\int (2 \pi)^{-6} d\mathbf{k}_1 d\mathbf{k}_2
\langle \mathbf{k}_1+\mathbf{a}_1/2,\mathbf{k}_2+\mathbf{a}_2/2
\vert \hat{v} \vert \mathbf{k}_1-\mathbf{a}_1/2,\mathbf{k}_2-\mathbf{a}_2/2\rangle
\times
\]
\[
e^{-i(  \mathbf{b}_1 \cdot (\mathbf{k}_1-\mathbf{a}_1/2)+
\mathbf{b}_2 \cdot (\mathbf{k}_2-\mathbf{a}_2/2)} .
\]
The Weyl kernels in (\ref{eq:tbc14}) become
\[
v(\mathbf{a}_1,(\mathbf{p}_2'-\mathbf{p}_2);\mathbf{b}_1,\mathbf{b}_2) =
\]
\[
\int (2 \pi)^{-6} d\mathbf{k}_1 d\mathbf{k}_2
\langle \mathbf{k}_1+\mathbf{a}_1/2,\mathbf{k}_2+(\mathbf{p}_2'-\mathbf{p}_2)/2
\vert \hat{v} \vert \mathbf{k}_1-\mathbf{a}_1/2,\mathbf{k}_2-(\mathbf{p}_2'-\mathbf{p}_2)/2\rangle \times
\]
\[
e^{-i(  \mathbf{b}_1 \cdot (\mathbf{k}_1-\mathbf{a}_1/2)+
\mathbf{b}_2 \cdot (\mathbf{k}_2-(\mathbf{p}_2'-\mathbf{p}_2)/2)} 
\]
\[
v((\mathbf{p}_1'-\mathbf{p}_1), \mathbf{a}_2,
;\mathbf{b}_1,\mathbf{b}_2) =
\]
\[
\int (2 \pi)^{-6} d\mathbf{k}_1 d\mathbf{k}_2
\langle \mathbf{k}_1+\mathbf{a}_1/2,\mathbf{k}_2+(\mathbf{p}_1'-\mathbf{p}_1)
/2
\vert \hat{v} \vert \mathbf{k}_1-(\mathbf{p}_1'-\mathbf{p}_1)
/2,\mathbf{k}_2-\mathbf{a}_2/2\rangle \times
\]
\beq
e^{-i(  \mathbf{b}_1 \cdot (\mathbf{k}_1-(\mathbf{p}_1'-\mathbf{p}_1)
/2)+
\mathbf{b}_2 \cdot (\mathbf{k}_2-\mathbf{a}_2/2)}.
\label{eq:tbc17}
\eeq
Inserting these expressions in the expression for the current matrix
elements gives
\[
\langle \mathbf{p}_2',\mathbf{p}'_1 \vert \mathbf{J}(\mathbf{x},0)
\vert \mathbf{p}_1 ,\mathbf{p}_2 \rangle =
\]
\[
-{ie_1 \over (2\pi)^9} \int d\mathbf{a}_1d\mathbf{b}_1d\mathbf{b}_2
 \int_0^1d\lambda \times
\]
\[
\int  d\mathbf{k}_1 d\mathbf{k}_2
\langle \mathbf{k}_1+\mathbf{a}_1/2,\mathbf{k}_2+(\mathbf{p}_2'-\mathbf{p}_2)/2 
\vert \hat{v} \vert \mathbf{k}_1-\mathbf{a}_1/2,\mathbf{k}_2-(\mathbf{p}_2'-\mathbf{p}_2)/2\rangle \times
\]
\[
e^{-i(  \mathbf{b}_1 \cdot (\mathbf{k}_1-\mathbf{a}_1/2)+
\mathbf{b}_2 \cdot (\mathbf{k}_2-(\mathbf{p}_2'-\mathbf{p}_2)/2)}
\times
\]
\[
\mathbf{b}_1 e^{i \mathbf{x}\cdot (\mathbf{a}_1+ \mathbf{p}_1
  -\mathbf{p}_1')} e^{i \mathbf{b}_2 \cdot \mathbf{p}_2 } e^{i
  \mathbf{b}_1 \cdot ( -\lambda \mathbf{a}_1-\lambda
  \mathbf{p}_1+\lambda \mathbf{p}_1' +\mathbf{p}_1)}
\]
\[
-{ie_2 \over (2\pi)^9} \int
d\mathbf{a}_2d\mathbf{b}_1d\mathbf{b}_2 \int_0^1d\lambda \times
\]
\[
\int  d\mathbf{k}_1 d\mathbf{k}_2
\langle \mathbf{k}_1+(\mathbf{p}_1'-\mathbf{p}_1)
/2 ,\mathbf{k}_2+\mathbf{a}_2/2 
\vert \hat{v} \vert \mathbf{k}_1-(\mathbf{p}_1'-\mathbf{p}_1)
/2,\mathbf{k}_2-\mathbf{a}_2/2\rangle \times
\]
\[
e^{-i(  \mathbf{b}_1 \cdot (\mathbf{k}_1-(\mathbf{p}_1'-\mathbf{p}_1)
/2)+
\mathbf{b}_2 \cdot (\mathbf{k}_2-\mathbf{a}_2/2)} \times
\]
\beq
\mathbf{b}_2 e^{i \mathbf{x}\cdot (\mathbf{a}_2+ \mathbf{p}_2
  -\mathbf{p}_2')} e^{i \mathbf{b}_1 \cdot \mathbf{p}_1 } e^{i
  \mathbf{b}_2 \cdot ( -\lambda \mathbf{a}_2-\lambda
  \mathbf{p}_2+\lambda \mathbf{p}_2' +\mathbf{p}_2)}.
\label{eq:tbc19a}
\eeq
Collecting terms in the exponents gives
\[
\langle \mathbf{p}_2',\mathbf{p}'_1 \vert \mathbf{J}(\mathbf{x},0)
\vert \mathbf{p}_1 ,\mathbf{p}_2 \rangle 
=-{i e_1\over (2\pi)^9} \int d\mathbf{a}_1d\mathbf{b}_1d\mathbf{b}_2
 \int_0^1d\lambda \times
\]
\[
\int  d\mathbf{k}_1 d\mathbf{k}_2
\langle \mathbf{k}_1+\mathbf{a}_1/2,\mathbf{k}_2+(\mathbf{p}_2'-\mathbf{p}_2)/2,
\vert \hat{v} \vert \mathbf{k}_1-\mathbf{a}_1/2,\mathbf{k}_2-(\mathbf{p}_2'-\mathbf{p}_2)/2\rangle \times
\]
\[
\mathbf{b}_1 e^{i \mathbf{b}_1 \cdot(-\mathbf{k}_1+\mathbf{a}_1/2
-\lambda \mathbf{a}_1-\lambda
\mathbf{p}_1+\lambda \mathbf{p}_1' +\mathbf{p}_1)  }
e^{i \mathbf{b}_2 \cdot( -\mathbf{k}_2+(1/2)(\mathbf{p}_2'-\mathbf{p}_2) +\mathbf{p}_2 )  }
e^{i \mathbf{x}  \cdot (\mathbf{a}_1+ \mathbf{p}_1
-\mathbf{p}_1')  }
\]
 \[
-{ie_2 \over (2\pi)^9} \int
d\mathbf{a}_2d\mathbf{b}_1d\mathbf{b}_2 \int_0^1d\lambda \times
\]
\[
\int  d\mathbf{k}_1 d\mathbf{k}_2
\langle \mathbf{k}_1+(\mathbf{p}_1'-\mathbf{p}_1)
/2 ,\mathbf{k}_2+\mathbf{a}_2/2
\vert \hat{v} \vert \mathbf{k}_1-(\mathbf{p}_1'-\mathbf{p}_1)
/2,\mathbf{k}_2-\mathbf{a}_2/2\rangle \times
\]
\beq
\mathbf{b}_2
e^{i \mathbf{b}_1\cdot (-\mathbf{k}_1+ (1/2)(\mathbf{p}_1'-\mathbf{p}_1)+ \mathbf{p}_1)}
e^{i \mathbf{b}_2\cdot (-\mathbf{k}_2+(1/2)\mathbf{a}_2
 -\lambda \mathbf{a}_2-\lambda
  \mathbf{p}_2+\lambda \mathbf{p}_2' +\mathbf{p}_2)}
e^{i \mathbf{x}\cdot (\mathbf{a}_2+ \mathbf{p}_2
  -\mathbf{p}_2')} .
\label{eq:tbc20a}
\eeq
Replacing
$\mathbf{b}_1=i \pmb{\nabla}_{k_1}$ and  
$\mathbf{b}_2= i \pmb{\nabla}_{k_2}$ acting on the exponent and
integrating over $\mathbf{b}_1$ and $\mathbf{b}_2$ gives $(2 \pi)^6$
times delta functions:
\[
\langle \mathbf{p}_2',\mathbf{p}'_1 \vert \mathbf{J}(\mathbf{x},0)
\vert \mathbf{p}_1 ,\mathbf{p}_2 \rangle
=-{i e_1\over (2\pi)^3} \int d\mathbf{a}_1 d\mathbf{k}_1 d\mathbf{k}_2
 \int_0^1d\lambda \times
\]
\[
\langle  \mathbf{k}_1+\mathbf{a}_1/2, \mathbf{k}_2+(\mathbf{p}_2'-\mathbf{p}_2)/2
\vert \hat{v} \vert \mathbf{k}_1-\mathbf{a}_1/2,\mathbf{k}_2-(\mathbf{p}_2'-\mathbf{p}_2)/2\rangle \times
\]
\[
i \pmb{\nabla}_{k_1}
\delta (-\mathbf{k}_1+\mathbf{a}_1/2
-\lambda \mathbf{a}_1-\lambda
\mathbf{p}_1+\lambda \mathbf{p}_1' +\mathbf{p}_1)  
\delta ( -\mathbf{k}_2+(1/2)(\mathbf{p}_2'-\mathbf{p}_2) +\mathbf{p}_2 )  \times
\]
\[e^{i \mathbf{x} \cdot(\mathbf{a}_1+ \mathbf{p}_1
-\mathbf{p}_1')  }
\]
\[
-{i e_2\over (2\pi)^3} \int
d\mathbf{a}_2 d\mathbf{k}_1 d\mathbf{k}_2 \int_0^1d\lambda \times
\]
\[
\langle \mathbf{k}_1+(\mathbf{p}_1'-\mathbf{p}_1)
/2,\mathbf{k}_2+\mathbf{a}_2/2 
\vert \hat{v} \vert \mathbf{k}_1-(\mathbf{p}_1'-\mathbf{p}_1)
/2,\mathbf{k}_2-\mathbf{a}_2/2\rangle \times
\]
\[
i 
\delta (-\mathbf{k}_1+ (1/2)(\mathbf{p}_1'-\mathbf{p}_1)+ \mathbf{p}_1)
\pmb{\nabla}_{k_2} \delta (-\mathbf{k}_2+(1/2)\mathbf{a}_2
-\lambda \mathbf{a}_2-\lambda
\mathbf{p}_2+\lambda \mathbf{p}_2' +\mathbf{p}_2) \times
\]
\beq
e^{i \mathbf{x} \cdot (\mathbf{a}_2+ \mathbf{p}_2
-\mathbf{p}_2')}.
\label{eq:tbc21a}
\eeq
Integrating the $\pmb{\nabla}_{k_i}$'s by parts gives
\[
\langle \mathbf{p}_2',\mathbf{p}'_1 \vert \mathbf{J}(\mathbf{x},0)
\vert \mathbf{p}_1 ,\mathbf{p}_2 \rangle  
=-{ e_1\over (2\pi)^3} \int d\mathbf{a}_1 d\mathbf{k}_1 d\mathbf{k}_2
 \int_0^1d\lambda \times
\]
\[
\pmb{\nabla}_{k_1} \langle  \mathbf{k}_1+\mathbf{a}_1/2, \mathbf{k}_2+(\mathbf{p}_2'-\mathbf{p}_2)/2
\vert \hat{v} \vert \mathbf{k}_1-\mathbf{a}_1/2,\mathbf{k}_2-(\mathbf{p}_2'-\mathbf{p}_2)/2\rangle \times
\]
\[
\delta (-\mathbf{k}_1+\mathbf{a}_1/2
-\lambda \mathbf{a}_1-\lambda
\mathbf{p}_1+\lambda \mathbf{p}_1' +\mathbf{p}_1)  
\delta ( -\mathbf{k}_2+(1/2)(\mathbf{p}_2'-\mathbf{p}_2) +\mathbf{p}_2 )  \times
\]
\[
e^{i \mathbf{x} \cdot(\mathbf{a}_1+ \mathbf{p}_1
-\mathbf{p}_1')  }
\]
\[
-{ e_2\over (2\pi)^3} \int
d\mathbf{a}_2 d\mathbf{k}_1 d\mathbf{k}_2 \int_0^1d\lambda \times
\]
\[
\pmb{\nabla}_{k_2} \langle \mathbf{k}_1+(\mathbf{p}_1'-\mathbf{p}_1)
/2,\mathbf{k}_2+\mathbf{a}_2/2 
\vert V \vert \mathbf{k}_1-(\mathbf{p}_1'-\mathbf{p}_1)
/2,\mathbf{k}_2-\mathbf{a}_2/2\rangle \times
\]
\[
\delta (-\mathbf{k}_1+ (1/2)(\mathbf{p}_1'-\mathbf{p}_1)+ \mathbf{p}_1)
\delta (-\mathbf{k}_2+(1/2)\mathbf{a}_2
-\lambda \mathbf{a}_2-\lambda
\mathbf{p}_2+\lambda \mathbf{p}_2' +\mathbf{p}_2) \times
\]
\beq
e^{i \mathbf{x} \cdot (\mathbf{a}_2+ \mathbf{p}_2
-\mathbf{p}_2')}.
\label{eq:tbc21aa}
\eeq
Now it is possible to perform both $\mathbf{k}$ integrals:
\[
\langle \mathbf{p}_2',\mathbf{p}'_1 \vert \mathbf{J}(\mathbf{x},0)
\vert \mathbf{p}_1 ,\mathbf{p}_2 \rangle 
=-{e_1 \over (2\pi)^3} \int d\mathbf{a}_1 
 \int_0^1d\lambda \times
\]
\[
\pmb{\nabla}_{11'} \langle
(1-\lambda)(\mathbf{p}_1+\mathbf{a}_1) +\lambda \mathbf{p}'_1,
\mathbf{p}_2' \vert \hat{v} \vert 
\lambda (\mathbf{p}_1' - \mathbf{a}_1)+(1-\lambda)\mathbf{p}_1 , \mathbf{p}_2
\rangle
e^{i \mathbf{x} \cdot(\mathbf{a}_1+ \mathbf{p}_1
-\mathbf{p}_1')  }
\]
\[
-{e_2 \over (2\pi)^3} \int
d\mathbf{a}_2  \int_0^1d\lambda \times
\]
\beq
\pmb{\nabla}_{22'}
\langle \mathbf{p}_1',(1-\lambda)(\mathbf{p}_2 + \mathbf{a}_2)+\lambda
\mathbf{p}_2' \vert \hat{v} \vert \mathbf{p}_1, \lambda (\mathbf{p}_2'- \mathbf{a_2)}
+ (1-\lambda) \mathbf{p}_2 \rangle
e^{i \mathbf{x} \cdot (\mathbf{a}_2+ \mathbf{p}_2
-\mathbf{p}_2')}
\label{eq:tbc21aaa}
\eeq
where in these expressions
\[
\nabla_{ii'}\langle \mathbf{p}_1,\mathbf{p}_2\vert \hat{v} \vert 
\mathbf{p}_1',\mathbf{p}_2' \rangle :=
\nabla_{p_i}\langle \mathbf{p}_1,\mathbf{p}_2\vert \hat{v} \vert
\mathbf{p}_1',\mathbf{p}_2' \rangle +
\nabla_{p_i'}\langle \mathbf{p}_1,\mathbf{p}_2\vert \hat{v} \vert
\mathbf{p}_1',\mathbf{p}_2' \rangle  .
\]
 
For $\mathbf{x}=0$ the current matrix elements become
\[
\langle \mathbf{p}_2',\mathbf{p}'_1 \vert \mathbf{J}(\mathbf{0},0)
\vert \mathbf{p}_1 ,\mathbf{p}_2 \rangle :=
\]
\[
=-{e_1 \over (2\pi)^3} \int d\mathbf{a}_1 
 \int_0^1d\lambda \times
\]
\[
\pmb{\nabla}_{11'} \langle
(1-\lambda)(\mathbf{p}_1+\mathbf{a}_1) +\lambda \mathbf{p}'_1,
\mathbf{p}_2' \vert \hat{v} \vert 
\lambda (\mathbf{p}_1' - \mathbf{a}_1)+(1-\lambda)\mathbf{p}_1 , \mathbf{p}_2
\rangle 
\]
\[
-{e_2 \over (2\pi)^3} \int
d\mathbf{a}_2  \int_0^1d\lambda \times
\]
\beq
\pmb{\nabla}_{22'}
\langle \mathbf{p}_1',(1-\lambda)(\mathbf{p}_2 + \mathbf{a}_2)+\lambda
\mathbf{p}_2' \vert \hat{v} \vert \mathbf{p}_1, \lambda (\mathbf{p}_2'- \mathbf{a}_2+
(1-\lambda) \mathbf{p}_2 \rangle.
\label{eq:tbc21aaaa}
\eeq
For interactions that conserve momentum the $\pmb{\nabla}_{ii'}$
it is useful to write the interaction in the form
\[
\langle \mathbf{p}'_1, \mathbf{p}'_2 \vert \hat{v} \vert \mathbf{p}_1, \mathbf{p}_2
\rangle = 
\delta (\mathbf{p}'_1 + \mathbf{p}'_2 - \mathbf{p}_1- \mathbf{p}_2 )
F_1 (\mathbf{p}_1',\mathbf{p}'_2,\mathbf{p}_2) +
\delta (\mathbf{p}'_1 + \mathbf{p}'_2 - \mathbf{p}_1- \mathbf{p}_2 )
F_2 (\mathbf{p}_2',\mathbf{p}'_1,\mathbf{p}_1) .
\]
Then
\[
\nabla_{ii}\langle \mathbf{p}_1,\mathbf{p}_2\vert \hat{v} \vert 
\mathbf{p}_1',\mathbf{p}_2' \rangle =
\delta (\mathbf{p}'_1 + \mathbf{p}'_2 - \mathbf{p}_1- \mathbf{p}_2 )
\nabla'_{i}F_i (\mathbf{p}_i',\mathbf{p}'_j,\mathbf{p}_j) \qquad j\not=i .
\]
Define
\[
\mathbf{F}_i (\mathbf{p}_i',\mathbf{p}'_j,\mathbf{p}_j) :=
\nabla'_{i}F_i (\mathbf{p}_i',\mathbf{p}'_j,\mathbf{p}_j)  \qquad j\not=i
\]
Using this in equation (\ref{eq:tbc21aaaa}) note that  
in the first term the momentum conserving delta function gives
$\mathbf{a}_1=\mathbf{p}'_2-\mathbf{p}_2$ while in the second term  
it gives $\mathbf{a}_2=\mathbf{p}'_1-\mathbf{p}_1$.  Since
$\pmb{\nabla}_{ii'}$ gives $0$ when acting on the momentum
conserving delta functions, it is possible to use the delta functions to
perform the $\mathbf{a}$ integrals.
\[
\langle \mathbf{p}_2',\mathbf{p}'_1 \vert \mathbf{J}(\mathbf{0},0)
\vert \mathbf{p}_1 ,\mathbf{p}_2 \rangle :=
\]
\[
=-{1 \over (2\pi)^3}  
 \int_0^1d\lambda 
\left (e_1 \mathbf{F}_1 ((1-\lambda)(\mathbf{p}_1+\mathbf{p}_2-\mathbf{p}_2') +\lambda \mathbf{p}'_1,
\mathbf{p}_2',\mathbf{p}_2) \right .
\]
\beq
\left .
+
e_2\mathbf{F}_2 ((1-\lambda)(\mathbf{p}_1+\mathbf{p}_2-\mathbf{p}_1') +\lambda \mathbf{p}'_2,
\mathbf{p}_1',\mathbf{p}_1) \right ).
\label{eq:tbc21ab}
\eeq

In the one-photon exchange approximation the two-body current
appears in the Hamiltonian in the form
\beq
\int d\mathbf{x}  \mathbf{A}(\mathbf{x},0)
\langle \mathbf{p}_1',\mathbf{p}'_2 \vert \mathbf{J}(\mathbf{x},0)
\vert \mathbf{p}_1 ,\mathbf{p}_2 \rangle =
(2 \pi)^{3/2} \tilde{\mathbf{A}}(\mathbf{Q},0)
\langle \mathbf{p}_2',\mathbf{p}'_1 \vert \mathbf{J}(\mathbf{0},0)
\vert \mathbf{p}_1 ,\mathbf{p}_2 \rangle
\label{eq:tbc30aa}
\eeq
where $\tilde{\mathbf{A}}(\mathbf{Q},0)$is the Fourier transform of the
vector potential at $t=0$:
\[
\tilde{\mathbf{A}}(\mathbf{Q},0) + {1 \over (2\pi)^{3/2}}\int d\mathbf{x} 
\mathbf{A}(\mathbf{x},0)e^{i \mathbf{Q}\cdot \mathbf{x}}
\]

\bigskip
\section{One-body currents}
\bigskip

A useful test is to apply the construction of the previous section to
a relativistic kinetic energy operator.
The relativistic kinetic energy for two relativistic particles has the form
\beq
\sqrt{\mathbf{p}_1^2 + m_2^2} + \sqrt{\mathbf{p}_2^2 + m_2^2}.  
\label{eq:ob0}
\eeq

If the momentum operators are replaced by covariant derivatives, then
the expression for the kinetic energy involves square roots of
sums of non-commuting operators.  One-body currents can be derived
using minimal substitution in the Weyl representation following
the construction used for the two-body interactions.
Because the momentum operators appear in the square roots,
the resulting current is no longer the familiar charge $\times$ velocity.

For each particle the method used to derive the two-body currents can
be applied to each one-body kinetic energy operator:
\[
 e{d \over de} 
\langle \mathbf{q} \vert e^{i\mathbf{b}\cdot (\mathbf{p}-e
\mathbf{A}(\mathbf{q}))} \vert \mathbf{p} \rangle_{e=0} = 
\langle \mathbf{q} \vert
\int_0^1 d\lambda e^{i\lambda \mathbf{b}\cdot \mathbf{p}} 
(-ie\mathbf{A(\mathbf{q})}\cdot \mathbf{b})  e^{i(1-\lambda) \mathbf{b}\cdot \mathbf{p}}
\vert \mathbf{p} \rangle =
\]
\[
\langle \mathbf{q} \vert
\int_0^1 d\lambda 
(-ie\mathbf{A(\mathbf{q}+\lambda \mathbf{b})}\cdot \mathbf{b})  e^{i \mathbf{b}\cdot \mathbf{p}}
\vert \mathbf{p} \rangle =
\]
\[
{1 \over (2 \pi)^{3/2} }
\int_0^1 d\lambda 
(-ie\mathbf{A(\mathbf{q}+\lambda \mathbf{b})}\cdot \mathbf{b})
e^{i (\mathbf{b}+\mathbf{q})\cdot \mathbf{p}} = 
\]
\[
{1 \over (2 \pi)^{3/2} }
\int d\mathbf{x} (-ie \mathbf{A}(\mathbf{x})\cdot
  \mathbf{b}) 
\int_0^1 d\lambda \delta (\mathbf{x} -\mathbf{q}- \lambda \mathbf{b}) e^{i (\mathbf{b}+\mathbf{q})\cdot \mathbf{p}} =
\]
\beq
{1 \over (2 \pi)^{3/2} }
\int d\mathbf{x} (-ie\mathbf{A}(\mathbf{x})\cdot \mathbf{b}) 
\int_0^1  d\lambda e^{i (
\mathbf{x}+(1- \lambda) \mathbf{b} )\cdot \mathbf{p}} .
\label{eq:ob1}
\eeq
Expression (\ref{eq:ob1}) can be used with an operator that is pure multiplication
in $\mathbf{p}$:
\beq
o(\mathbf{a},\mathbf{b}) = {1 \over (2 \pi)^3}
\delta(\mathbf{a}) \int d\mathbf{k} e^{-i \mathbf{b}\cdot \mathbf{k}}\sqrt{\mathbf{k}^{2}+ {m}^2}.
\label{eq:ob2}
\eeq
Using this in the expression for the gauge covariant single particle
energy gives
\[
\langle \mathbf{q} \vert \sqrt{ (\mathbf{p}- e\mathbf{A})^2 + m^2}
\vert \mathbf{p}\rangle =
\]
\[
{1 \over (2 \pi)^3} \int d\mathbf{k}d\mathbf{b}d\mathbf{x} e^{-i \mathbf{b}\cdot \mathbf{k}}\sqrt{\mathbf{k}^{2}+ \mathbf{m}^2}
{1 \over (2 \pi)^{3/2} } (-ie\mathbf{A}(\mathbf{x})\cdot
\mathbf{b})  \times
\]
\[
\int_0^1 d\lambda \delta (\mathbf{x} -\mathbf{q}- \lambda \mathbf{b}) e^{i (
\mathbf{x}+(1- \lambda) \mathbf{b} )\cdot \mathbf{p}} =
\]

\beq
\int d\mathbf{k}d\mathbf{b}d\mathbf{x} {1 \over (2 \pi)^{9/2}} (-ie\mathbf{A}(\mathbf{x})\cdot
\mathbf{b})
\int_0^1 d\lambda \delta (\mathbf{x} -\mathbf{q}- \lambda \mathbf{b}) \sqrt{\mathbf{k}^{ 2}+ \mathbf{m}^2}
e^{i (
\mathbf{x}+(1- \lambda) \mathbf{b} )\cdot \mathbf{p} - \mathbf{b}\cdot \mathbf{k}}).
\label{eq:ob3}
\eeq
Next transform from a mixed representation to a momentum representation
\[
e{d \over de}
\langle \mathbf{p}' \vert \sqrt{ (\mathbf{p}- e\mathbf{A})^2 + m^2}
\vert \mathbf{p}\rangle =
\]
\[ 
\int d\mathbf{x}d\mathbf{k} d\mathbf{q} d\mathbf{b}
{1 \over (2 \pi)^{6}} (-ie\mathbf{A}(\mathbf{x})\cdot
\mathbf{b})
\int_0^1 d\lambda \delta (\mathbf{x} -\mathbf{q}- \lambda \mathbf{b}) \sqrt{\mathbf{k}^{2}+ \mathbf{m}^2}
e^{i (
  \mathbf{x}+(1- \lambda) \mathbf{b} )\cdot \mathbf{p} - \mathbf{k}\cdot \mathbf{b}
- \mathbf{p}' \cdot \mathbf{q})
}
=
\]
\[ 
\int d\mathbf{x} d\mathbf{k} d\mathbf{b}
{1 \over (2 \pi)^{6}}  (-ie\mathbf{A}(\mathbf{x})\cdot
\mathbf{b})
\int_0^1 d\lambda  \sqrt{\mathbf{k}^{2}+ \mathbf{m}^2}
e^{i (
  \mathbf{x}+(1- \lambda) \mathbf{b} )\cdot \mathbf{p} - \mathbf{k}\cdot \mathbf{b}
- \mathbf{p}' \cdot (\mathbf{x}- \lambda \mathbf{b}))
}
=
\]
\beq
\int d\mathbf{x} d\mathbf{k} d\mathbf{b}
{1 \over (2 \pi)^{6}} (-ie\mathbf{A}(\mathbf{x})\cdot \mathbf{b})
\int_0^1 d\lambda  \sqrt{\mathbf{k}^{2}+ \mathbf{m}^2}
e^{i(\mathbf{x} \cdot (\mathbf{p}-\mathbf{p}')+
\mathbf{b} \cdot ((1-\lambda)\mathbf{p}-\mathbf{k}+ \lambda \mathbf{p}')) }
.
\label{eq:ob4}
\eeq
After replacing $\mathbf{b}$ by $i\pmb{\nabla}_k$ acting on the exponent 
the $\mathbf{b}$ integral can be performed which gives
\beq
(i\pmb{\nabla}_k) (2 \pi)^3 \delta ((1-\lambda)\mathbf{p}-\mathbf{k}+ \lambda \mathbf{p}').
\label{eq:ob5}
\eeq
After this the $\mathbf{k}$ integral can be evaluated giving
\beq
\mathbf{k} =(1-\lambda)\mathbf{p} + \lambda \mathbf{p}'
\label{eq:ob6}
\eeq
and after integrating by parts the term linear in the vector
potential becomes
\beq 
-\int d\mathbf{x} e\mathbf{A}(\mathbf{x})
{1 \over (2 \pi)^{3}} 
\int_0^1 d\lambda \cdot
{ (1-\lambda)\mathbf{p} + \lambda \mathbf{p}' \over 
\sqrt{((1-\lambda)\mathbf{p} + \lambda \mathbf{p}')^2+ \mathbf{m}^2}}
e^{i\mathbf{x} \cdot (\mathbf{p}-\mathbf{p}') }.
\label{eq:ob7}
\eeq
This gives a one-body current matrix element of the form
\beq
\langle \mathbf{p}' \vert \mathbf{J}(\mathbf{x},0) \vert \mathbf{p} \rangle=
- {1 \over (2 \pi)^{3}} 
\int_0^1 d\lambda 
{(1-\lambda) \mathbf{p} + \lambda \mathbf{p}') \over 
\sqrt{((1-\lambda) \mathbf{p} + \lambda \mathbf{p}')^2+ \mathbf{m}^2}}
e^{i\mathbf{x} \cdot (\mathbf{p}-\mathbf{p}') }.
\label{eq:ob8}
\eeq
If the $\mathbf{x}$ integral in (\ref{eq:ob7}) is used to Fourier transform
the vector potential the current for two non-interacting relativistic particles becomes 
\[
\langle \mathbf{p}_1',\mathbf{p}_2' \vert \mathbf{J}(\mathbf{0},0) \vert
\mathbf{p}_1',\mathbf{p}_2' 
\rangle =
-{1 \over (2 \pi)^{3/2}} 
\int_0^1 d\lambda
\]
\[
\left [e_1
{ (1-\lambda) \mathbf{p}_1 + \lambda \mathbf{p}_1') \over 
  \sqrt{((1-\lambda) \mathbf{p}_1 + \lambda \mathbf{p}_1')^2+ \mathbf{m}_1^2}}
\delta(\mathbf{p}_2'-\mathbf{p}_2)+
\right .
\]
\beq
\left .
e_2
{(1-\lambda) \mathbf{p}_2 + \lambda \mathbf{p}_2') \over 
  \sqrt{((1-\lambda)\mathbf{p}_2 + \lambda \mathbf{p}_2')^2+ \mathbf{m}_2^2}}
\delta(\mathbf{p}_1'-\mathbf{p}_1)
\right ].
\label{eq:ob9}
\eeq
The non-trivial convolution arises because of the square root factor.

It is instructive to compare this to the non-relativistic case.
In the non-relativistic case $\sqrt{\mathbf{k}^2 + m^2}$ is replaced by
$\mathbf{k}^2/2m$.  In that case equation (\ref{eq:ob9}) becomes
\[
\langle \mathbf{p}_1',\mathbf{p}_2' \vert \mathbf{J}(\mathbf{0},0)
\vert \mathbf{p}_1,\mathbf{p}_2 \rangle=
-{1 \over (2 \pi)^{3/2}}  
\int_0^1 d\lambda 
    [e_1{(1-\lambda) \mathbf{p}_1 + \lambda \mathbf{p}_1') \over 2m_1}
    \delta(\mathbf{p}_2'-\mathbf{p}_2)
    \]
    \[
    +
  {e_2(1-\lambda) \mathbf{p}_2 + \lambda \mathbf{p}_2') \over 2 m_2}
  \delta(\mathbf{p}_1'-\mathbf{p}_1)] =
    \]
    \[
-{1 \over (2 \pi)^{3/2}}  
   [e_1 {(\mathbf{p}_1 +\mathbf{p}_1') \over 2m_1}
    \delta(\mathbf{p}_2'-\mathbf{p}_2)
    \]
    \beq
    +
  e_2 {(\mathbf{p}_2 + \mathbf{p}_2') \over 2 m_2}
  \delta(\mathbf{p}_1'-\mathbf{p}_1)] .
\label{eq:ob10}
\eeq
Both (\ref{eq:ob10}) and
(\ref{eq:ob9}) have the form of charge $\times$ velocity
where the velocities involve the average of the initial and final
relativistic respectively non-relativistic velocities.
  
As mentioned earlier, this results in a convection current for a
point charge.  If the relativistic kinetic energy (\ref{eq:ob0}) is
replaced by the sum of one-body Dirac Hamiltonians, the resulting one
body-currents will have both a convection and magnetic component,
however they will not have nucleon form factors.  This is because nucleons
are composite systems with non-trivial internal charge and current
distributions.

\section{Charge density}

Replacing the momentum operators in the Weyl representation of the interaction
by covariant derivatives results in a vector current, but it does not result in
the full four current.

Since the 4-current transforms the same way as the four momentum under
Lorentz transformations, current covariance can be used to compute the
charge density operator in terms of the vector part of the current.
The most straightforward way to determine the charge density from the
vector part or the current is use the commutator 
with the dynamical rotationless boost generators
\beq
J^0(\mathbf{q},0) =  i[K^i,J^i(\mathbf{q},0)] 
\qquad \mbox{no sum, any }i,
\label{eq:cd1}
\eeq
where $\mathbf{K}$ is the generator of rotationless boosts.
This has the advantage that the result is an operator (not a matrix element)
that transforms as a 4 vector under the original representation of the
Poincar\'e group.  This is consistent with the one-photon exchange
approximation.  

Both the boost generators and the Hamiltonian depend on the interactions.
If the momenta in the expression for all of the generators were replaced by
covariant derivatives, the resulting operators would no longer satisfy the
Poincar\'e commutation relations.  This is because, unlike ordinary
partial derivatives, different components of the covariant derivatives
do not commute.  However any locally gauge invariant extension
of the full Poincare Lie algebra should reduce to the original algebra
in the limit of zero charge.  The current is the coefficient of the term
in the gauge invariant Hamiltonian that is linear in the vector potential
so the transformation properties of the current in the one photon exchange
approximation is determined by the original representation of the
Poincar\'e Lie algebra.  This assumes that
it is possible to construct locally gauge invariant boost that is consistent
with the gauge invariant Hamiltonian.  

To
compute $J^0(\mathbf{q},0)$ it is necessary to have explicit dynamical
boost generators.  The two-body Bakamjian-Thomas construction \cite{Bakamjian:1953kh} leads to
an explicit expression for the dynamical boost generators as a function of
the dynamical Hamiltonian and non-interacting one-body generators.  
Formally the Bakamjian-Thomas dynamical boost generator is 
\beq
\mathbf{K}={1 \over 2} (H\mathbf{X}_0+ \mathbf{X}_0H)-
{\mathbf{s}_0 \times\mathbf{P}_0
\over M+H}  
\label{eq:cd2}
\eeq
where $\mathbf{s}_0$ is the non-interacting two-body canonical spin
and $\mathbf{X}_0$ is the Newton-Wigner position operator for the
non-interacting two-body system.  In this case all of the interaction
dependence appears in $M$ and $H$.  Expressions for boost generators in
representations that satisfy many-body cluster separability
can be constructed following
\cite{Sokolov:1977}\cite{Coester:1982vt}.

The non-interacting two-body $\mathbf{X}_0$ and $\mathbf{s}_0$ operators
are the
following functions of the non-interacting one-body Poincar\'e generators 
\[
\mathbf{X}_0 =-\frac{1}{2} \{ {1 \over H_1+H_2},\mathbf{K}_1+\mathbf{K}_2\}
\]
\beq
- { (\mathbf{P}_1 + \mathbf{P}_2)\times \left (
(H_1+H_2)  (\mathbf{J}_1 + \mathbf{J}_2)-   
(\mathbf{P}_1 + \mathbf{P}_2) \times (\mathbf{K}_1 + \mathbf{K}_2)
\right )  
\over   ((H_1+H_2)M_0(H_1+H_2+M_0) }
\label{eq:cd3}
\eeq
where
\beq
\mathbf{s}_0=(\mathbf{J}_1 + \mathbf{J}_2) - \mathbf{X}_0\times
(\mathbf{P}_1 + \mathbf{P}_2)
\label{eq:cd4}
\eeq
\beq
M_0 = \sqrt{ (H_1+H_2)^2 - (\mathbf{P}_1 + \mathbf{P}_2)^2}.
\label{eq:cd5}
\eeq
\beq
\mathbf{K}_i =
{1 \over 2} \{\sqrt{\mathbf{p}_i^2 + m_i^2}, (i\pmb{\nabla}_{p_1}\}-
{\mathbf{s}_i \times\mathbf{p}_i
\over m_i + \sqrt{\mathbf{p}_i^2 + m_i^2}}. 
\label{eq:cd6}
\eeq
In these expressions $\mathbf{J}$ is an angular momentum generator
rather than a current operator.
The advantage is that
these are operator definitions that can be consistently applied to
different kinds of reactions.
While these are complex expressions at the
operator level, they are easy to compute in matrix elements
between irreducible eigenstates, where $\mathbf{s}_0$ is the spin of
the initial or final two-body state, $\mathbf{X}_0$ is (i $\times$ )
the partial derivative with respect to the total linear momentum
holding the spin constant, and $H$ and $M$ are the energy and mass
eigenvalues of the initial or final state.  Since
\[
e^{i K_x \rho} \vert (m,j)\mathbf{P},\mu \rangle =
\vert (m,j)\pmb{\Lambda}_x(\rho){P},\nu \rangle
D^j_{\mu \nu}[B^{-1} (\Lambda_x^{-1}(\rho) p)\Lambda_x(\rho)B(p) ]
\]
where ${\Lambda}_x(\rho)$ is a rotationless boost in the $x$ direction
with rapidity $\rho$ and $B(p)$ is a rotationless boost
that transforms a particle at rest to one moving with momentum $\mathbf{p}$.
Tt follows that   
\[
K_x\vert (m,j)\mathbf{P},\mu \rangle =
-i \frac{\partial }{ \partial \rho} 
\vert (m,j)\pmb{\Lambda}_x(\rho){P},\nu \rangle
D^j_{\mu \nu}[B^{-1} (\Lambda_x^{-1}(\rho) p)\Lambda_x(\rho)B(p) ]_{\rho=0}
\]
can be obtained by differentiating the Lorentz boosted state
by the rapidity at zero rapidity.

For systems of more than two particles the Bakamjian-Thomas
construction does not lead to Poincar\'e generators that satisfy cluster
properties.  This can be repaired \cite{Sokolov:1977}\cite{Coester:1982vt} at the
expense of introducing many-body interactions in the Hamiltonian,
boost generators and spin operators.

\section{Time derivatives}

The fields in the Hamiltonian are at a fixed common time, $t$.
The gauge transformations discussed in the previous section were 
spatial gauge transformations at a single fixed time.  
The Schr\"odinger equation involves time derivatives.
To make the time-dependent
Schr\"odinger
equation invariant under gauge transformations
the time derivative also has to be replaced by a covariant derivative.

Gauge invariance can be achieved by replacing the time derivative by 
\beq
(\partial_t -i \sum_i e_i A_0 (\mathbf{q}_i,t)).
\label{eq:td1}
\eeq
To see this note that under a local gauge transformation 
\beq
\psi (\mathbf{q}_1, \cdots , \mathbf{q}_N ,t) \to
\label{eq:td2}
\eeq
\beq
e^{i \sum \chi (\mathbf{q}_i,t) }\psi (\mathbf{q}_1, \cdots , \mathbf{q}_N ,t). 
\label{eq:td3}
\eeq
The time derivative of the transformed wave function is
\beq
e^{i \sum \phi (\mathbf{q}_i,t) }
(i\partial_t - \sum_i \partial_t \chi (\mathbf{q}_i,t))
\psi (\mathbf{q}_1, \cdots , \mathbf{q}_N ,t).
\label{eq:td4}
\eeq
The derivatives of the phases can be eliminated by replacing the
time derivative by
\beq
(i \partial_t + \sum_i e_iA_0 (\mathbf{q}_i,t)) \Psi (\mathbf{q}_1, \cdots , \mathbf{q}_N ,t).
\label{eq:td5}
\eeq
In this way the gauge invariant Schr\"odinger equation becomes
\[
i \partial_t \Psi (\mathbf{q}_1, \cdots , \mathbf{q}_N ,t) =
\]
\beq
- \sum_i eA_0 (\mathbf{q}_i,t) \Psi (\mathbf{q}_1, \cdots , \mathbf{q}_N ,t)
+H(\mathbf{p}-e\mathbf{A},\mathbf{q}) \Psi (\mathbf{q}_1, \cdots , \mathbf{q}_N ,t)
\label{eq:td6}
\eeq
where the gauge invariant Hamiltonian is constructed by replacing
$\mathbf{p}_i$ by $\mathbf{p}_i - e_i\mathbf{A}(\mathbf{q}_i)$ in the
Weyl representation of the Hamiltonian.

This prescription results in a locally gauge covariant equation that
reduces to the ordinary relativistic Schr\"odinger equation in the
limit that the charge vanishes.

\section{Spin-dependent interactions} 

\bigskip

The canonical pairs of operators
used in Weyl representation were the single-particle momentum and the
single-particle Newton-Wigner position operators, which in the
momentum-canonical spin basis are $i \pmb{\nabla}_{\mathbf{p}_i}$ where
the partial derivatives are computed holding the single particle
canonical spins constant.
In this representation, since the derivatives commute with the spin,
the gauge transformations are independent of the canonical spin.  This
is not entirely trivial since the spins undergo momentum dependent
Wigner rotations under boosts.
The procedure is still the same.  The first step is to express
the reduced interaction in terms of single-nucleon momenta and spins.
Since the single-particle canonical spins are gauge invariant in this
representation, the current has a structure similar to (\ref{eq:tbc21aaa})
where the potential kernel has both single-particle momenta and spins:

For interactions that conserve momentum the
interactions can be expressed in the form:
\[
\langle \mathbf{p}'_1,\mu_1', \mathbf{p}'_2, \mu_2',\vert \hat{v} \vert \mathbf{p}_1,\mu_1, \mathbf{p}_2,\mu_2
\rangle =
\]
\[
\delta (\mathbf{p}'_1 + \mathbf{p}'_2 - \mathbf{p}_1- \mathbf{p}_2 )
(F_1 (\mathbf{p}_1',\mathbf{p}'_2,\mathbf{p}_2, \mu_1',\mu_2',\mu_1,\mu_2) 
+
F_2 (\mathbf{p}_2',\mathbf{p}'_1,\mathbf{p}_1,\mu_1',\mu_2',\mu_1,\mu_2) .
\]
The current is expressed in terms of the derivatives
\[
\nabla_{ii}
\langle \mathbf{p}'_1,\mu_1', \mathbf{p}'_2, \mu_2',\vert \hat{v} \vert \mathbf{p}_1,\mu_1, \mathbf{p}_2,\mu_2
\rangle  =
\]
\[
\delta (\mathbf{p}'_1 + \mathbf{p}'_2 - \mathbf{p}_1- \mathbf{p}_2 )
\nabla'_{i}F_i (\mathbf{p}_i',\mathbf{p}'_j,\mathbf{p}_j,\mu_1',\mu_2',\mu_1,\mu_2) \qquad j\not=i .
\]
which are used to define
\[
\mathbf{F}_i (\mathbf{p}_i',\mathbf{p}'_j,\mathbf{p}_j,\mu_1',\mu_2',\mu_1,\mu_2) :=
\nabla'_{i}F_i (\mathbf{p}_i',\mathbf{p}'_j,\mathbf{p}_j,\mu_1',\mu_2',\mu_1,\mu_2)  \qquad j\not=i .
\]

Using this in equation (\ref{eq:tbc21aaaa}) note that  
in the first term the momentum conserving delta function gives
$\mathbf{a}_1=\mathbf{p}'_2-\mathbf{p}_2$ while in the second term  
it gives $\mathbf{a}_2=\mathbf{p}'_1-\mathbf{p}_1$.  Since
$\pmb{\nabla}_{ii'}$ gives $0$ when acting on the momentum
conserving delta functions, it is possible to use the delta functions to
perform the $\mathbf{a}$ integrals.
\[
\langle \mathbf{p}_2',\mathbf{p}'_1 \vert \mathbf{J}(\mathbf{0},0)
\vert \mathbf{p}_1 ,\mathbf{p}_2 \rangle :=
\]
\[
=-\frac{1}{(2\pi)^3}  
 \int_0^1d\lambda \times
\]
\beq
\left ( \mathbf{F}_1 ((1-\lambda)(\mathbf{p}_1+\mathbf{p}_2-\mathbf{p}_2') +\lambda \mathbf{p}'_1,
\mathbf{p}_2',\mathbf{p}_2)
+
\mathbf{F}_2 ((1-\lambda)(\mathbf{p}_1+\mathbf{p}_2-\mathbf{p}_1') +\lambda \mathbf{p}'_2,
\mathbf{p}_1',\mathbf{p}_1) \right ).
\label{eq:tbc21xxx}
\eeq

If the interaction is expressed in terms of kinematic masses and
kinematically invariant degeneracy parameters,
\beq
\delta (\mathbf{P}' - \mathbf{P})\delta_{j'j}\delta_{\mu'\mu}
\langle k',l' ,s' \Vert \hat{v}^j \Vert k,l,s \rangle 
\label{eq:s4}
\eeq
then the Clebsch-Gordon
coefficients for the Poincar\'e group
\cite{Keister:1991sb}\cite{Polyzou:2002id} need to be used to get the
equivalent expression in terms of single-particle spins and momenta that can be differentiated

The Clebsch-Gordon coefficients for the Poincar\'e group that relate
the non-interacting irreducible representation of the Poincar\'e group in
the momentum-canonical spin basis to the tensor product of two
single-particle irreducible representations are
\[
\vert \mathbf{p}_1,\mu_1, \mathbf{p}_2, \mu_2, \rangle =
\vert
\mathbf{P},\mu, l, s, k(p_1,p_2) \rangle
\langle s,\mu_s,l,m \vert j, \mu \rangle \times
\]
\[
\sqrt{  \frac{(\omega_{1}(\mathbf{p}_1)+\omega_{2}(\mathbf{p}_1))
\omega_{1}(\mathbf{k}) \omega_{2}(\mathbf{k})}{
(\omega_{1}(\mathbf{k})+\omega_{2}(\mathbf{k}) )\omega_{1}(\mathbf{p}_1)
\omega_{2}(\mathbf{p}_2)}}\times
\]
\beq
\langle \frac{1}{2}, \mu_1'', \frac{1}{2}, \mu_2'' \vert S, \mu_s  \rangle
Y^{l*}_m (\hat{\mathbf{k}}(p_1,p_2))
D^{1/2}_{\mu_1'' \mu_1}(R_w(P,p_1))
D^{1/2}_{\mu_2'' \mu_2}(R_w(P,p_2)),
\label{eq:s5}
\eeq
where
\beq
\mathbf{P}= \mathbf{p}_1 + \mathbf{p}_2 \qquad
\omega_i (\mathbf{p}_i) = \sqrt{m_i^2 + \mathbf{p}_i^2},
\label{eq:s6}
\eeq
\[
\mathbf{k}={1 \over 2}(\mathbf{k}_1-\mathbf{k}_2)
\]
\[
\mathbf{k}_i = \mathbf{p}_i
\frac{\mathbf{P}}{\sqrt{
(\omega_1 (\mathbf{p}_1) +\omega_2 (\mathbf{p}_2))^2-\mathbf{P}^2}}
\times
\]
\beq
\left (
\frac{(\mathbf{p}_1^2 - \mathbf{p}^2_2)}{
\sqrt{
(\omega_1 (\mathbf{p}_1) +\omega_2 (\mathbf{p}_2))^2-\mathbf{P}^2}+
\omega_1 (\mathbf{p}_1) +\omega_2 (\mathbf{p}_2)
} - \omega_i (\mathbf{p}_i) 
\right )
\eeq
\beq
H_0= \omega_{m_1} (\mathbf{p}_1) + \omega_{m_2} (\mathbf{p}_2),
\qquad
M_0= \sqrt{H_0^2 -\mathbf{P}^2} =  \omega_{m_1} (\mathbf{k}_1) + \omega_{m_2} (\mathbf{k}_2),
\label{eq:s7}
\eeq
$R_w(P,p_2)$ is the Wigner rotation
\beq
R_w(P,p_i) = B^{-1}(\mathbf{k}_i/m_i) B^{-1} (\mathbf{P}/M_0) B(\mathbf{p}_i/m_i),
\label{eq:s8}
\eeq
\beq
k_i = B^{-1} (\mathbf{P}/M_0) p_i = (\sqrt{\mathbf{k}_i^2+m^2},\mathbf{k}_i)
\label{eq:s9}
\eeq
\beq
\mathbf{k}= {1 \over 2} (\mathbf{k}_1-\mathbf{k}_2)
\label{eq:s10}
\eeq
and $B(\mathbf{p}/m)$ is a rotationless Lorentz boost:
\beq
B^{-1} (\mathbf{p}/m) = B^{-1} (-\mathbf{p}/m)
\label{eq:s11}
\eeq
\beq
B(\mathbf{p}/m)^{0}{}_{0} = {\omega_{m} (\mathbf{p}) \over m}
\qquad
B(\mathbf{p}/m)^{0}{}_i = B(\mathbf{p}/m)^{i}{}_0 = {p^i\over m}
\label{eq:s12}
\eeq
\beq
B(\mathbf{p}/m)^{i}{}_j = \delta^{ij} + \frac{p^ip^j}{ m(\omega_{m} (\mathbf{p}))}.
\label{eq:s13}
\eeq

Light-front formulations of relativistic quantum mechanics are useful
for studying hadronic structure because the boosts are kinematic.  The
general method discussed in this paper can also be extended to the
light-front generators.

The light-front Poincare generators are different linear combinations
of the Poincar\'e infinitesimal generators where seven of the new
generators can be chosen to have no interactions.

The light front Hamiltonian is
\[
P^- = H-P^3 .
\]
An irreducible set of canonical operators can be expressed in terms of
kinematic generators
\[
P^{+} = H + P^3, \qquad\mbox{and} \qquad  \mathbf{E}_{\perp} = \mathbf{K}_{\perp}- \hat{\mathbf{z}}\times \mathbf{J} .
\]
The canonical pairs are
\[
p_i^+, q_i^- = \frac{1}{ 2} \{ \frac{1}{ p_i^+},K_i^3\} = i \frac{\partial }{ \partial p_i^+}
\]
\[
\mathbf{p_{i\perp}}, \mathbf{q_{i\perp}} = i \frac{1 }{ p^+}\mathbf{E}_{\perp}=
 i \frac{\partial }{ \partial p_{\perp i}}
\]
and the Weyl representation for the interacting $P^-$ operator is
\[
\hat{P}^-= \int d^{3N}\mathbf{a} d^{3N}\mathbf{b} {p^-(\mathbf{a},\mathbf{b})}
e^{i \mathbf{a}\cdot \hat{\mathbf{q}}} e^{i \mathbf{b}\cdot {\hat{\mathbf{p}}}}
\]
The current can be extracted using same methods used in sections
seven and eight.  In this case the current will be different because the
partial derivatives are computed holding the light-front spin constant.
The light front and canonical spins are related by momentum dependent
Melosh \cite{Melosh:1974cu} rotations.
  
\section{Summary - Conclusion}

In this work current operators in phenomenological relativistic models
of strongly interacting systems were constructed by requiring local
gauge invariance of the Hamiltonian.  In these models impulse
(one-body currents) are not compatible with the dynamical constraints
of current covariance and current conservation.  These constraints
alone are insufficient to uniquely fix a current operator.  Local gauge
invariance was implemented by expressing the Hamiltonian in the Weyl
representation, replacing the single-particle momentum operators by
gauge covariant derivatives.  The dynamical current was identified with
the coefficient of the part of gauge invariant Hamiltonian that is linear
in the vector potential.  The charge density was constructed using the
commutation relations with the dynamical Lorentz boost generators, which
ensures that the 4-current transforms as a 4-vector density.

The advantage of this construction is that the result is a current
operator that can be used in reactions with different initial and
final states that is consistent with the dynamics.  The operators in
(\ref{ex:7}), (\ref{eq:tbc21aaa}) and (\ref{eq:ob2}) have an explicit dependence on the
representation of the Hamiltonian.  The construction addresses one of
the primary challenges in constructing relativistic models, which is
the absence of a systematic method for constructing candidates for
dynamical current operators that are consistent with the interaction.
While the result is not unique, it is a 4-vector current operator that
in minimally consistent with the interaction.

A relativistic Hamiltonian theory with a finite number of degrees of
freedom can be formulated on a many-particle Hilbert space.  In the
absence of interactions particles transform under irreducible
representations of the Poincar\'e group.  The momentum and coordinates
in the Weyl representation are the single-particle momentum and
Newton-Wigner position operators.  The Newton-Wigner position
operator, which in the momentum-canonical spin basis, is $i\, \times$
the partial derivative with respect to the momentum holding the single
particle canonical spin constant.  Since the dynamical relativistic
model is formulated on the same many-body Hilbert space and the Weyl
algebra is irreducible, the Hamiltonian and all of the Poincar\'e
generators can be represented in the free-particle Weyl
representation.

The dynamical models under consideration are defined by a dynamical
unitary representation of the Poincar\'e group acting on the many
particle Hilbert space \cite{Sokolov:1977ym}\cite{Coester:1982vt}\cite{Polyzou:2003dt}.  In this application
the dynamical representation is chosen so translations and rotations
are independent of interactions (Dirac's instant form of dynamics
\cite{Dirac:1949cp}).
Solving the dynamics is equivalent to decomposing this unitary
representation into a direct integral of irreducible representations.

The Hamiltonians in these models are typically non-local. Gauge
invariance requires that the kernel of the Hamiltonian in the ``position
representation'' satisfies the intertwining properties (\ref{eq:d2}-\ref{eq:d3}), which
follow as a consequence of replacing the momentum operator in the Weyl
representation by covariant derivatives.  The analysis involves
evaluating functions of non-commuting variables, but the final
results are explicit expressions for the kernels.

Examples of the currents that come from non-relativistic
$\mathbf{L}\cdot \mathbf{S}$, $(\mathbf{L}\cdot \mathbf{S})$ and
$\mathbf{L}\cdot \mathbf{L}$ interactions were given.  The resulting
currents are explicit
function of the interactions.

While this construction is systematic and yields a covariant current
operator that is consistent with the dynamics, whether the dynamical
mechanism generates currents consistent with experiment needs further
investigation.
One problem that was not addressed is by replacing the momentum
operators by covariant derivatives, the Poincar\'e commutation
relation are no longer preserved, since the commutator of two
covariant derivatives is the field strength tensor rather than 0.
Fortunately in the one-photon exchange approximation this problem goes
away when the higher powers of the vector potential are set to 0.
Going beyond the one-photon exchange approximation requires further
investigation.

This research was supported by the US Department of Energy, Office of Science, 
grant number DE-SC0016457.

\section{Appendix}

Expressions (\ref{ex:8}) and (\ref{ex:9}) for the two-body currents
associated with the operators $(\mathbf{L}\cdot \mathbf{S})^2$ and $(\mathbf{L}\cdot
\mathbf{L})$ in non-relativistic interactions are derived in this
appendix.  The construction of these current operators is similar to
the construction used in the $\mathbf{L}\cdot \mathbf{S}$ case.  The main
difference is that these contributions involve products of covariant
derivatives.  The current is the coefficient of the vector potential.

The interactions before replacing the derivatives by
covariant derivatives are
\[
v_{(\mathbf{L}\cdot \mathbf{S})^2}(\vert \mathbf{q} \vert)
(\mathbf{L}\cdot \mathbf{S})^2 =
v_{(\mathbf{L}\cdot \mathbf{S})^2}(\vert \mathbf{q} \vert)
(\mathbf{S} \times \mathbf{q})\cdot \mathbf{p} 
(\mathbf{S} \times \mathbf{q})\cdot \mathbf{p} =
\]
\beq
v_{(\mathbf{L}\cdot \mathbf{S})^2}(\vert \mathbf{q} \vert)[
\sum_{ij}(\mathbf{S} \times \mathbf{q})_i 
(\mathbf{S} \times \mathbf{q})_j p_ip_j
+i (\mathbf{S} \times \mathbf{q}) \cdot
(\mathbf{S} \times \mathbf{p})]
\label{ap:8}
\eeq
and
\beq
v_{\mathbf{L}\cdot \mathbf{L}}(\vert \mathbf{q} \vert)
\mathbf{L}\cdot \mathbf{L} =
v_{\mathbf{L}\cdot \mathbf{L}}(\vert \mathbf{q} \vert)[
\mathbf{q}^2 \mathbf{p}^2 - \mathbf{q} \cdot (\mathbf{q} \cdot \mathbf{p}) \mathbf{p} + 2 i \mathbf{q} \cdot \mathbf{p}].
\label{ap:9}
\eeq
where in these expressions the canonical commutation relations are used to
move the momentum operators to the right of the coordinate operators.

In order to extract the current it is useful to express (\ref{ap:8})
and (\ref{ap:9}) in terms of single-particle variables, where minimal
substitution is straightforward:
\[
v_{(\mathbf{L}\cdot \mathbf{S})^2}(\vert \mathbf{q} \vert)
(\mathbf{L}\cdot \mathbf{S})^2 =
\]
\[
\sum_{ij}v_{(\mathbf{L}\cdot \mathbf{S})^2}(\vert \mathbf{q}_1-\mathbf{q}_2 \vert)[
(\mathbf{S} \times (\mathbf{q}_1-\mathbf{q}_2))_i 
  (\mathbf{S} \times (\mathbf{q}_1-\mathbf{q}_2))_j \frac{(\mathbf{p}_1 - \mathbf{p}_2)_i }{ 2}\frac{(\mathbf{p}_1 - \mathbf{p}_2)_j }{ 2}
\]
\beq
+\frac{i}{ 2} (\mathbf{S} \times (\mathbf{q}_1- \mathbf{q}_2)) \cdot
(\mathbf{S} \times (\mathbf{p}_1 -\mathbf{p}_2))]
\label{ap:10}
\eeq
and
\[
v_{\mathbf{L}\cdot \mathbf{L}}(\vert \mathbf{q}_1-\mathbf{q}_2 \vert)
\mathbf{L}\cdot \mathbf{L} =
\]
\[
v_{\mathbf{L}\cdot \mathbf{L}}(\vert \mathbf{q}_1-\mathbf{q}_2 \vert)[
  {1 \over 4}(\mathbf{q}_1-\mathbf{q}_2)^2 (\mathbf{p}_1-\mathbf{p}_2)^2
  \]
  \beq
- {1 \over 4}(\mathbf{q}_1-\mathbf{q}_2) \cdot ((\mathbf{q}_1-\mathbf{q}_2)
\cdot (\mathbf{p}_1-\mathbf{p}_2))( \mathbf{p}_1-\mathbf{p}_2) +
 i (\mathbf{q}_1 -\mathbf{q}_2) \cdot (\mathbf{p}_1- \mathbf{p}_2)].
\label{ap:11}
\eeq
These are operator expressions.  Replacing the space derivatives by covariant
derivatives is equivalent to replacing
$\mathbf{p}_i$ by $\mathbf{p}_i -e_i \mathbf{A}(\mathbf{q}_i)$,
maintaining the correct operator ordering.  This substitution results in the
gauge invariant operators
\[
v_{(\mathbf{L}\cdot \mathbf{S})^2}(\vert \mathbf{q} \vert)
(\mathbf{L}\cdot \mathbf{S})^2 \to
\]
\[
v_{(\mathbf{L}\cdot \mathbf{S})^2}(\vert \mathbf{q}_1-\mathbf{q}_2 \vert)
\left [\frac{1 }{ 4} \sum_{ij}
(\mathbf{S} \times (\mathbf{q}_1-\mathbf{q}_2))_i 
(\mathbf{S} \times (\mathbf{q}_1-\mathbf{q}_2))_j \times
\right .
\]
\[
\left .
(\mathbf{p}_1 -e_1 \mathbf{A}(\mathbf{q}_1) - \mathbf{p}_2 +e_2 \mathbf{A}(\mathbf{q}_2))_i
  (\mathbf{p}_1 -e_1 \mathbf{A}(\mathbf{q}_1) - \mathbf{p}_2 +e_2 \mathbf{A}(\mathbf{q}_2))_j + \right .
\]
\beq
\left .
\frac{i}{ 2} (\mathbf{S} \times (\mathbf{q}_1- \mathbf{q}_2)) \cdot
  (\mathbf{S} \times
{(\mathbf{p}_1 -e_1 \mathbf{A}(\mathbf{q}_1) - \mathbf{p}_2 +e_2 \mathbf{A}(\mathbf{q}_2))} \right ]
\label{ap:12}
\eeq
and
\[
v_{\mathbf{L}\cdot \mathbf{L}}(\vert \mathbf{q}_1-\mathbf{q}_2 \vert)
\mathbf{L}\cdot \mathbf{L} \to
v_{\mathbf{L}\cdot \mathbf{L}}(\vert \mathbf{q}_1-\mathbf{q}_2 \vert)
\times
\]
\[
\left [
\frac{1}{4}(\mathbf{q}_1-\mathbf{q}_2)^2
{(\mathbf{p}_1 -e_1 \mathbf{A}(\mathbf{q}_1) - \mathbf{p}_2 +e_2 \mathbf{A}(\mathbf{q}_2))}\cdot
{(\mathbf{p}_1 -e_1 \mathbf{A}(\mathbf{q}_1) - \mathbf{p}_2 +e_2 \mathbf{A}(\mathbf{q}_2))} - \right .
\]
\[
- \frac{1}{4}(\mathbf{q}_1-\mathbf{q}_2) \cdot [(\mathbf{q}_1-\mathbf{q}_2)
\cdot
{(\mathbf{p}_1 -e_1 \mathbf{A}(\mathbf{q}_1) - \mathbf{p}_2 +e_2 \mathbf{A}(\mathbf{q}_2)) }
\cdot 
{(\mathbf{p}_1 -e_1 \mathbf{A}(\mathbf{q}_1) - \mathbf{p}_2 +e_2 \mathbf{A}(\mathbf{q}_2)) }
\]
\beq
\left .
+ i (\mathbf{q}_1 -\mathbf{q}_2) \cdot
{(\mathbf{p}_1 -e_1 \mathbf{A}(\mathbf{q}_1) - \mathbf{p}_2 +e_2 \mathbf{A}(\mathbf{q}_2))}
\right ] .
\label{ap:13}
\eeq
Keeping only the terms that are linear in $\mathbf{A}$
in (\ref{ap:12}) and (\ref{ap:13}) and expressing the result in terms of the
total and relative momenta and their conjugate coordinates gives:
\[
(\ref{ap:12})\to v_{(\mathbf{L}\cdot \mathbf{S})^2}(\vert \mathbf{q}_1-\mathbf{q}_2 \vert)
\times
\]
\[
\left [-\frac{1 }{ 2} 
(\mathbf{S} \times \mathbf{q})_i 
(\mathbf{S} \times \mathbf{q})_j
\{ \mathbf{p}_i
(e_1\mathbf{A}(\mathbf{Q}+\frac{\mathbf{q}}{ 2}) -e_2\mathbf{A}((\mathbf{Q}-
   \frac{\mathbf{q}}{ 2}))_j +\right .
   \]
\[
\left .
(e_1\mathbf{A}(\mathbf{Q}+\frac{\mathbf{q}}{ 2}) -e_2\mathbf{A}(\mathbf{Q}
-\frac{\mathbf{q}}{ 2}))_i \mathbf{p}_j\}
  + \right .
\]
\beq
\left .
-\frac{i}{ 2} (\mathbf{S} \times \mathbf{q}) \cdot
(\mathbf{S} \times
(e_1\mathbf{A}(\mathbf{Q}+\frac{\mathbf{q}}{2}) - e_2\mathbf{A}
(\mathbf{Q}-\frac{\mathbf{q}}{ 2})) \right ]
\label{ap:14}
\eeq
and
\[
(\ref{ap:13})\to  v_{\mathbf{L}\cdot \mathbf{L}}(\vert \mathbf{q} \vert)\left \{
-\frac{1}{2}\mathbf{q}^2 \left [
(\mathbf{p}\cdot
( e_1\mathbf{A}(\mathbf{Q}+\frac{\mathbf{q}}{ 2}) -
  e_2\mathbf{A}(\mathbf{Q}-\frac{\mathbf{q}}{ 2})) +
  \right . \right .
  \]
  \[
  \left .
 \left . 
(e_1\mathbf{A}(\mathbf{Q}+\frac{\mathbf{q}}{2}) -
e_2\mathbf{A}(\mathbf{Q}-\frac{\mathbf{q}}{ 2}))
\cdot \mathbf{p}) \right ]
\right . +
\]
\[
+ \frac{1}{ 2}\{ \mathbf{q} \cdot (\mathbf{q}
\cdot \mathbf{p}) 
\cdot 
(e_1\mathbf{A}(\mathbf{Q}+\frac{\mathbf{q}}{2})- e_2\mathbf{A}(\mathbf{Q}-
    \frac{\mathbf{q}}{ 2}))
+ \frac{1}{ 2}\mathbf{q} \cdot (\mathbf{q}
\cdot
(e_1\mathbf{A}(\mathbf{Q}+\frac{\mathbf{q}}{ 2})- e_2\mathbf{A}(\mathbf{Q}-
    \frac{\mathbf{q}}{ 2}) \mathbf{p} \}
\]
\beq
\left .\left .
-i  \mathbf{q}  \cdot
( e_1\mathbf{A}(\mathbf{Q}+\frac{\mathbf{q}}{ 2}) -e_2 \mathbf{A}(\mathbf{Q}-
\frac{\mathbf{q}}{ 2})
\right ] \right \}
\label{ap:15}
\eeq
where in (\ref{ap:14}) $\mathbf{p}_i$ represents the $i$-th component of
the relative momentum rather than the momentum of particle $i$
The next step is to use the commutation relations to move the $\mathbf{p}$
factors to the right of all of the coordinate factors.   There are
three terms in (\ref{ap:14}) and (\ref{ap:15}) where the momenta are
on the left of some of the coordinate operators.  These terms are:
\[
\mathbf{p}_i
(e_1\mathbf{A}(\mathbf{Q}+\frac{\mathbf{q}}{ 2}) -e_2\mathbf{A}((\mathbf{Q}-
       \frac{\mathbf{q}}{ 2}))_j =
\]
\beq
(e_1\mathbf{A}(\mathbf{Q}+\frac{\mathbf{q}}{ 2}) -e_2\mathbf{A}((\mathbf{Q}-
    \frac{\mathbf{q}}{ 2}))_j \mathbf{p}_i - \frac{i}{ 2}  
\partial_{Qi} (e_1\mathbf{A}(\mathbf{Q}+\frac{\mathbf{q}}{ 2}) +e_2\mathbf{A}(\mathbf{Q}-\frac{\mathbf{q}}{ 2}))_j
\label{ap:16}
\eeq

\bigskip
 
\[
-\frac{1}{ 2}\mathbf{q}^2
\mathbf{p}\cdot
(e_1\mathbf{A}(\mathbf{Q}+\frac{\mathbf{q}}{ 2}) -
e_2\mathbf{A}(\mathbf{Q}-\frac{\mathbf{q}}{ 2})) =
\]
\beq
-\frac{1}{ 2}\mathbf{q}^2
(e_1\mathbf{A}(\mathbf{Q}+\frac{\mathbf{q}}{ 2}) -
e_2\mathbf{A}(\mathbf{Q}-\frac{\mathbf{q}}{ 2}))
\cdot \mathbf{p}
+i\frac{1}{4}\mathbf{q}^2
\pmb{\nabla}_Q\cdot (e_1\mathbf{A}(\mathbf{Q}+\frac{\mathbf{q}}{ 2}) +
e_2\mathbf{A}(\mathbf{Q}-\frac{\mathbf{q}}{ 2})),
\label{ap:17}
\eeq

\bigskip

\[
\frac{1}{ 2}\{ \mathbf{q} \cdot (\mathbf{q}
\cdot \mathbf{p}) 
\cdot 
(e_1\mathbf{A}(\mathbf{Q}+\frac{\mathbf{q}}{ 2})- e_2\mathbf{A}(\mathbf{Q}-
    \frac{\mathbf{q}}{2}))=
\]
\beq
\frac{1}{ 2}\{ 
\mathbf{q} \cdot 
(e_1\mathbf{A}(\mathbf{Q}+\frac{\mathbf{q}}{ 2})- e_2\mathbf{A}(\mathbf{Q}-
       \frac{\mathbf{q}}{ 2})) (\mathbf{q}
\cdot \mathbf{p}) 
-i{1 }{4}\{ 
\mathbf{q} \cdot (\mathbf{q} \cdot \pmb{\nabla}_Q) 
(e_1\mathbf{A}(\mathbf{Q}+\frac{\mathbf{q}}{ 2})+ e_2\mathbf{A}(\mathbf{Q}-
       \frac{\mathbf{q}}{ 2})).  
\label{ap:18}
\eeq
Using these identities in the operator expressions
(\ref{ap:14}) and (\ref{ap:15})
above gives
\[
(\ref{ap:14})=
\]
\[
v_{(\mathbf{L}\cdot \mathbf{S})^2}(\vert \mathbf{q} \vert)
\left [-\frac{1}{ 2} 
(\mathbf{S} \times \mathbf{q})_i 
(\mathbf{S} \times \mathbf{q})_j
\{ 
(e_1\mathbf{A}(\mathbf{Q}+\frac{\mathbf{q}}{2}) -e_2\mathbf{A}(\mathbf{Q}-
  \frac{\mathbf{q}}{ 2}))_j \mathbf{p}_i + \right .
\]
\[
(e_1\mathbf{A}(\mathbf{Q}+\frac{\mathbf{q}}{2}) -e_2\mathbf{A}(\mathbf{Q}-
  \frac{\mathbf{q}}{2}))_i\mathbf{p}_j
\]
\[
- \frac{i }{ 2}  
((e_1\partial_{iQ} \mathbf{A}(\mathbf{Q}+\frac{\mathbf{q}}{ 2}) +e_2\partial_{iQ}\mathbf{A}(\mathbf{Q}-\frac{\mathbf{q}}{ 2}))_j \}
\]
\beq
\left .
-\frac{i}{ 2} (\mathbf{S} \times \mathbf{q}) \cdot
(\mathbf{S} \times
(e_1\mathbf{A}(\mathbf{Q}+\frac{\mathbf{q}}{2}) - e_2\mathbf{A}
(\mathbf{Q}-\frac{\mathbf{q}}{ 2})) \right ] 
\label{ap:19}
\eeq
and
\[
(\ref{ap:15})=
\]
\[
v_{\mathbf{L}\cdot \mathbf{L}}(\vert \mathbf{q} \vert)\left [
-\frac{1 }{2}\mathbf{q}^2
(
2 (e_1\mathbf{A}(\mathbf{Q}+\frac{\mathbf{q}}{2}) -
e_2\mathbf{A}(\mathbf{Q}-\frac{\mathbf{q}}{ 2}))
\cdot \mathbf{p}) \right .
\]
\[
+i\frac{1}{ 4}\mathbf{q}^2
(e_1\pmb{\nabla}\cdot \mathbf{A}(\mathbf{Q}+\frac{\mathbf{q}}{ 2})+
e_2\pmb{\nabla}\cdot \mathbf{A}(\mathbf{Q}-\frac{\mathbf{q}}{ 2}))
 +
\]
\[
+ \{ \mathbf{q}  
\cdot 
(e_1\mathbf{A}(\mathbf{Q}+\frac{\mathbf{q}}{2})- e_2\mathbf{A}(\mathbf{Q}-
    \frac{\mathbf{q}}{ 2}))(\mathbf{q}
\cdot \mathbf{p})
-i\frac{1}{4}\{ 
\mathbf{q} \cdot (\mathbf{q} \cdot \pmb{\nabla}_Q) 
(e_1\mathbf{A}(\mathbf{Q}+\frac{\mathbf{q}}{2})+ e_2\mathbf{A}(\mathbf{Q}-
       \frac{\mathbf{q}}{ 2}))  
\]
\beq
\left .
-i  \mathbf{q}  \cdot
( e_1\mathbf{A}(\mathbf{Q}+\frac{\mathbf{q}}{ 2}) - e_2\mathbf{A}(\mathbf{Q}-
\frac{\mathbf{q}}{ 2})
\right ].
\label{ap:20}
\eeq
Since the momentum operators are to
the right of the coordinate operators, the operators become numbers in
a mixed coordinate-momentum basis.  The mixed matrix elements
of (\ref{ap:19}) and (\ref{ap:20}) are
\[
\langle \mathbf{Q},\mathbf{q} \vert
(\ref{ap:19})\vert \mathbf{P},\mathbf{p} \rangle =
\]
\[
v_{(\mathbf{L}\cdot \mathbf{S})^2}(\vert \mathbf{q} \vert)
\frac{1 }{ (2\pi)^3} e^{i \mathbf{Q}\cdot \mathbf{P} + i \mathbf{q}\cdot \mathbf{p}}
\left [ - (\mathbf{S}\times \mathbf{q})_i (\mathbf{S}\times \mathbf{q})_j  \left \{
%
p_i (e_1\mathbf{A}(\mathbf{Q}+\frac{\mathbf{q}}{ 2}) -e_2\mathbf{A}((\mathbf{Q}-\frac{\mathbf{q}}{ 2}))_j   
 \right .\right .
\]
\[
\left .
\frac{i}{ 4}  
(e_1\partial_{iQ} \mathbf{A}(\mathbf{Q}+\frac{\mathbf{q}}{ 2}) +e_2\partial_{iQ}\mathbf{A}(\mathbf{Q}-\frac{\mathbf{q}}{ 2}))_j  \right \} 
\]
\beq
\left .
-\frac{i}{ 2} (\mathbf{S} \times \mathbf{q}) \cdot
(\mathbf{S} \times
(e_1\mathbf{A}(\mathbf{Q}+\frac{\mathbf{q}}{ 2}) - e_2\mathbf{A}
(\mathbf{Q}-\frac{\mathbf{q}}{ 2})) \right ]
\label{ap:21}
\eeq
and
\[
\langle \mathbf{Q},\mathbf{q} \vert 
(\ref{ap:20})\vert  \mathbf{P},\mathbf{p} \rangle =
\]
\[
v_{\mathbf{L}\cdot \mathbf{L}}(\vert \mathbf{q} \vert)
\frac{1 }{ (2\pi)^3} e^{i \mathbf{Q}\cdot \mathbf{P} + i \mathbf{q}\cdot \mathbf{p}}
\left [
-\mathbf{q}^2
( e_1\mathbf{A}(\mathbf{Q}+\frac{\mathbf{q}}{ 2} -
e_2 \mathbf{A}(\mathbf{Q}-\frac{\mathbf{q}}{ 2}))
\cdot \mathbf{p}
\right .
\]
\[
\left .
+\frac{i}{ 4}\mathbf{q}^2
(e_1\pmb{\nabla}_Q\cdot \mathbf{A}(\mathbf{Q}+\frac{\mathbf{q}}{ 2})+
e_2\pmb{\nabla}_Q\cdot \mathbf{A}(\mathbf{Q}-\frac{\mathbf{q}}{ 2}))
\right . +
\]
\[
+   (\mathbf{q}
\cdot \mathbf{p}) \mathbf{q} \cdot 
(e_1\mathbf{A}(\mathbf{Q}+\frac{\mathbf{q}}{2})- e_2\mathbf{A}(\mathbf{Q}-
      \frac{\mathbf{q}}{ 2}))
\]
\[
-\frac{i}{ 4}\{ 
\mathbf{q} \cdot (\mathbf{q} \cdot \pmb{\nabla}_Q) 
(e_1\mathbf{A}(\mathbf{Q}+\frac{\mathbf{q}}{ 2})+ e_2\mathbf{A}(\mathbf{Q}
-\frac{\mathbf{q}}{ 2}))  
\]
\beq
\left .
-i \mathbf{q}  \cdot
( e_1\mathbf{A}(\mathbf{Q}+\frac{\mathbf{q}}{ 2}) - e_2\mathbf{A}(\mathbf{Q}-
\frac{\mathbf{q}}{2})
\right ]
\label{ap:22}
\eeq
Next change the 
final variables to momentum variables.  With this change
these expressions become
\[
\langle \mathbf{P}',\mathbf{p}' \vert
\ref{ap:19}) 
\vert  \mathbf{P}',\mathbf{p}' \rangle =
\]
\[
\int d\mathbf{q} d \mathbf{Q}
v_{(\mathbf{L}\cdot \mathbf{S})^2}(\vert \mathbf{q} \vert)
\frac{1}{ (2\pi)^6} e^{i \mathbf{Q}\cdot (\mathbf{P}-\mathbf{P}') + i \mathbf{q}\cdot
  (\mathbf{p}-\mathbf{p}')} \times
\]
\[
\left [-
(\mathbf{S} \times \mathbf{q})\cdot\mathbf{p}  
(\mathbf{S} \times \mathbf{q}) \cdot
\{ 
(e_1\mathbf{A}(\mathbf{Q}+\frac{\mathbf{q}}{ 2}) -e_2\mathbf{A}((\mathbf{Q}-
  \frac{\mathbf{q}}{ 2}))
+ \right .
\]
-\[\sum_{ij}
\frac{i}{ 4}(\mathbf{S} \times \mathbf{q})_i (\mathbf{S} \times \mathbf{q})_j
(e_1\partial_i \mathbf{A}(\mathbf{Q}+\frac{\mathbf{q}}{ 2}) +e_2\partial_i\mathbf{A}(\mathbf{Q}-\frac{\mathbf{q}}{ 2}))_j 
\]
\beq
\left .
-\frac{i}{ 2}(\mathbf{S}^2 \mathbf{q}- (\mathbf{q}\cdot \mathbf{S}) \mathbf{S}
)\cdot 
(e_1\mathbf{A}(\mathbf{Q}+\frac{\mathbf{q}}{ 2}) - e_2\mathbf{A}
(\mathbf{Q}-\frac{\mathbf{q}}{ 2})) \right ].
\label{ap:23}
\eeq
and
\[
\langle \mathbf{P}',\mathbf{p}' \vert
(\ref{ap:20})
\vert  \mathbf{P}',\mathbf{p}' \rangle =
\]
\[
\int d\mathbf{q} d \mathbf{Q}
v_{\mathbf{L}\cdot \mathbf{L}}(\vert \mathbf{q} \vert)
\frac{1 }{ (2\pi)^6} e^{i \mathbf{Q}\cdot (\mathbf{P}-\mathbf{P}') + i \mathbf{q}\cdot
  (\mathbf{p}-\mathbf{p}')} \times
\]
\[
\left [
-\mathbf{q}^2
(\mathbf{p}\cdot 
(e_1\mathbf{A}(\mathbf{Q}+\frac{\mathbf{q}}{ 2}) -
e_2\mathbf{A}(\mathbf{Q}-\frac{\mathbf{q}}{ 2}))
 +\frac{i }{ 4}\mathbf{q}^2
(e_1\pmb{\nabla}_Q\cdot \mathbf{A}(\mathbf{Q}+\frac{\mathbf{q}}{ 2})+
e_2\pmb{\nabla}_Q\cdot \mathbf{A})\mathbf{Q}-\frac{\mathbf{q}}{ 2})
)\right . +
\]
\[
+ \{  \cdot (\mathbf{q}
\cdot \mathbf{p}) \mathbf{q} 
\cdot 
(e_1\mathbf{A}(\mathbf{Q}+\frac{\mathbf{q}}{ 2})- e_2\mathbf{A}(\mathbf{Q}-
    \frac{\mathbf{q}}{ 2}))
\frac{i}{4}\{ 
\mathbf{q} \cdot (\mathbf{q} \cdot \pmb{\nabla}) 
(e_1\mathbf{A}(\mathbf{Q}+\frac{\mathbf{q}}{ 2})+ e_2\mathbf{A}(\mathbf{Q}-
       \frac{\mathbf{q}}{ 2}))  
\]
\beq
\left .
-i \mathbf{q}  \cdot
( e_1 \mathbf{A}(\mathbf{Q}+\frac{\mathbf{q}}{ 2}) -e_2 \mathbf{A}(\mathbf{Q}-\frac{\mathbf{q}}{ 2})
\right ] 
\label{ap:24}
\eeq

In (\ref{ap:23}) and (\ref{ap:24}) the derivatives can be removed
from the vector potential by integrating by parts, assuming no contribution from the boundary terms.

The three terms in these expressions with derivatives on the vector potential  are
\[
\frac{i}{ 4}\frac{1}{ (2\pi)^6}\sum_{ij}
v_{(\mathbf{L}\cdot \mathbf{S})^2}(\vert \mathbf{q} \vert)
 e^{i \mathbf{Q}\cdot (\mathbf{P}-\mathbf{P}') + i \mathbf{q}\cdot
  (\mathbf{p}-\mathbf{p}')} 
(\mathbf{S} \times \mathbf{q})_i 
 (\mathbf{S} \times \mathbf{q})_j \times
 \]
 \beq
(e_1\partial_i \mathbf{A}(\mathbf{Q}+\frac{\mathbf{q}}{ 2}) +e_2\partial_i\mathbf{A}(\mathbf{Q}-\frac{\mathbf{q}}{ 2}))_j \}
\label{ap:25}
\eeq
\bigskip
\[
\frac{i }{ 4}\frac{1 }{ (2\pi)^6}
\int d\mathbf{q} d \mathbf{Q}
v_{\mathbf{L}\cdot \mathbf{L}}(\vert \mathbf{q} \vert)
e^{i \mathbf{Q}\cdot (\mathbf{P}-\mathbf{P}') + i \mathbf{q}\cdot
(\mathbf{p}-\mathbf{p}')}
\mathbf{q}^2 \times
\]
\beq
(e_1\pmb{\nabla}\cdot \mathbf{A}(\mathbf{Q}+\frac{\mathbf{q}}{ 2})+
e_2\pmb{\nabla}\cdot \mathbf{A}(\mathbf{Q}-\frac{\mathbf{q}}{ 2})
)
\label{ap:26}
\eeq
and
\bigskip
\[
-\frac{i}{ 4}\frac{1 }{ (2\pi)^6}
\int d\mathbf{q} d \mathbf{Q}
v_{\mathbf{L}\cdot \mathbf{L}}(\vert \mathbf{q} \vert)
e^{i \mathbf{Q}\cdot (\mathbf{P}-\mathbf{P}') + i \mathbf{q}\cdot
(\mathbf{p}-\mathbf{p}')}
\{ 
\mathbf{q} \cdot (\mathbf{q} \cdot \pmb{\nabla})
\]
\beq
(e_1\mathbf{A}(\mathbf{Q}+\frac{\mathbf{q}}{ 2})+ e_2\mathbf{A}(\mathbf{Q}-
    \frac{\mathbf{q}}{ 2})).  
\label{ap:27}
\eeq

The partial derivatives in (\ref{ap:25}-\ref{ap:27}) are derivatives of the argument of the vector potential, equivalently with respect to the $\mathbf{Q}$ variables.  They can be replaced by $\pm 2\partial_{q_i}$
\[
-\frac{ie}{ 2}\frac{1 }{ (2\pi)^6} \sum_{ij}
\int d\mathbf{q} d \mathbf{Q}
(\partial_{q_i} [
v_{(\mathbf{L}\cdot \mathbf{S})^2}(\vert \mathbf{q} \vert)
 e^{i \mathbf{Q}\cdot (\mathbf{P}-\mathbf{P}') + i \mathbf{q}\cdot
  (\mathbf{p}-\mathbf{p}')}
(\mathbf{S} \times \mathbf{q})_i 
 (\mathbf{S} \times \mathbf{q})_j] \times
\]
\beq
(e_1\mathbf{A}(\mathbf{Q}+\frac{\mathbf{q}}{ 2}) -
e_2\mathbf{A}(\mathbf{Q}-\frac{\mathbf{q}}{ 2}))_j 
\label{ap:28}
\eeq
\bigskip
\beq
-\frac{i}{ 2}\frac{1 }{ (2\pi)^6}
\int d\mathbf{q} d \mathbf{Q}
\pmb{\nabla}_q 
[
v_{\mathbf{L}\cdot \mathbf{L}}(\vert \mathbf{q} \vert)
\frac{1}{ (2\pi)^6} e^{i \mathbf{Q}\cdot (\mathbf{P}-\mathbf{P}') + i \mathbf{q}\cdot
  (\mathbf{p}-\mathbf{p}')}
\mathbf{q}^2 ]\cdot 
(e_1\mathbf{A}(\mathbf{Q}+\frac{\mathbf{q}}{ 2})-
e_2\mathbf{A}(\mathbf{Q}-\frac{\mathbf{q}}{ 2})
)
\label{ap:29}
\eeq
\bigskip
\beq
\frac{i}{ 2}\frac{1}{ (2\pi)^6}
\int d\mathbf{q} d \mathbf{Q}
 \pmb{\nabla}_q [\cdot \mathbf{q}
   v_{\mathbf{L}\cdot \mathbf{L}}(\vert \mathbf{q} \vert)
   e^{i \mathbf{Q}\cdot (\mathbf{P}-\mathbf{P}') + i \mathbf{q}\cdot
(\mathbf{p}-\mathbf{p}')}
\mathbf{q} ]\cdot
 (e_1\mathbf{A}(\mathbf{Q}-\frac{\mathbf{q}}{ 2})- e_2\mathbf{A}(\mathbf{Q}-
     \frac{\mathbf{q}}{ 2})) . 
\label{ap:30}
\eeq
Using (\ref{ap:28}-\ref{ap:30}) and the identities
\beq
(e_1\mathbf{A}(\mathbf{Q}+\frac{\mathbf{q}}{ 2})- e_2\mathbf{A}(\mathbf{Q}-
    \frac{\mathbf{q}}{ 2})) =
\int d\mathbf{x} 
(e_1\delta (\mathbf{x} -\mathbf{Q}-\frac{\mathbf{q}}{ 2})-
\delta (\mathbf{x} -e_2\mathbf{Q}+\frac{\mathbf{q}}{ 2}))
A(\mathbf{x},0)
\label{ap:31}
\eeq
and
\[
\int d\mathbf{Q}
e^{i \mathbf{Q}\cdot (\mathbf{P}-\mathbf{P}')}
(e_1\delta (\mathbf{x} -\mathbf{Q}-\frac{\mathbf{q}}{ 2})-
e_2\delta (\mathbf{x} -\mathbf{Q}+\frac{\mathbf{q}}{ 2})) =
%
\]
\[
e^{i \mathbf{x}\cdot (\mathbf{P}-\mathbf{P}')}(
e^{i \mathbf{q}\cdot (\mathbf{P}-\mathbf{P}')/2} - e^{-i \mathbf{q}\cdot (\mathbf{P}-\mathbf{P}')/2})=
\]
\beq
e^{i \mathbf{x}\cdot (\mathbf{P}-\mathbf{P}')}[
  (e_1-e_2) \cos \left (\frac{\mathbf{q}\cdot (\mathbf{P}'-\mathbf{P}}{ 2} \right) +i (e_1+e_2) 
\sin \left (\frac{\mathbf{q}\cdot (\mathbf{P}'-\mathbf{P}}{ 2} \right )]
\label{ap:33}
\eeq
in the expressions for the currents gives the result below for the
three derivative terms. 
Since the $\mathbf{q}$ in these expressions comes from the vector potential,
it does not get touched by derivatives.  Since there is no other $\mathbf{Q}$
-dependence these factors multiply everything. 
\[
-\frac{i}{4}\frac{1}{ (2\pi)^6}
\int d\mathbf{q} 
(\partial_{q_i} [
v_{(\mathbf{L}\cdot \mathbf{S})^2}(\vert \mathbf{q} \vert)
 e^{i \mathbf{x}\cdot (\mathbf{P}-\mathbf{P}') + i \mathbf{q}\cdot
  (\mathbf{p}-\mathbf{p}')} 
(\mathbf{S} \times \mathbf{q})_i 
 (\mathbf{S} \times \mathbf{q})] \times
\]
\beq
[(e_1-e_2) \cos \left (\frac{\mathbf{q}\cdot (\mathbf{P}'-\mathbf{P}}{ 2} \right) +i (e_1+e_2) 
\sin \left (\frac{\mathbf{q}\cdot (\mathbf{P}'-\mathbf{P}}{2} \right )]
\label{ap:34}
\eeq
\bigskip
\[
-\frac{i }{ 2}\frac{1}{ (2\pi)^6}
\int d\mathbf{q}
\pmb{\nabla}_q\cdot 
[
v_{\mathbf{L}\cdot \mathbf{L}}(\vert \mathbf{q} \vert)
e^{i \mathbf{x}\cdot (\mathbf{P}-\mathbf{P}') + i \mathbf{q}\cdot
(\mathbf{p}-\mathbf{p}')}
\mathbf{q}^2 ]
\times
\]
\beq
[ (e_1-e_2) \cos \left (\frac{\mathbf{q}\cdot (\mathbf{P}'-\mathbf{P}}{ 2} \right) +i (e_1+e_2) 
\sin \left (\frac{\mathbf{q}\cdot (\mathbf{P}'-\mathbf{P}}{ 2} \right )]
\label{ap:35}
\eeq
\bigskip
\[
\frac{i}{ 2} \frac{1}{ (2\pi)^6}
\int d\mathbf{q} 
 \pmb{\nabla}_{q_i}[\cdot \mathbf{q} 
   v_{\mathbf{L}\cdot \mathbf{L}}(\vert \mathbf{q} \vert)
   e^{i \mathbf{x}\cdot (\mathbf{P}-\mathbf{P}') + i \mathbf{q}\cdot
     (\mathbf{p}-\mathbf{p}')} \mathbf{q}] \times
\]
\beq
[
(e_1-e_2) \cos \left (\frac{\mathbf{q}\cdot (\mathbf{P}'-\mathbf{P}}{ 2} \right) +i (e_1+e_2) 
\sin \left (\frac{\mathbf{q}\cdot (\mathbf{P}'-\mathbf{P}}{ 2} \right )]
\label{ap:36}
\eeq
These derivative expressions go in the expression for the current
matrix elements,
which is the coefficient of the vector potential.
\[
\langle \mathbf{P}',\mathbf{p'} \vert \mathbf{J}_{(\mathbf{L}\cdot \mathbf{S})^2}(\mathbf{x},0)
\vert \mathbf{P},\mathbf{p} \rangle=
\int d\mathbf{q} 
v_{(\mathbf{L}\cdot \mathbf{S})^2}(\vert \mathbf{q} \vert)
{1 \over (2\pi)^6} e^{i \mathbf{x}\cdot (\mathbf{P}-\mathbf{P}') + i \mathbf{q}\cdot
  (\mathbf{p}-\mathbf{p}')} \times
\]
\[
[ (e_1-e_2) \cos \left (\frac{\mathbf{q}\cdot (\mathbf{P}'-\mathbf{P}}{ 2} \right) +i (e_1+e_2) 
\sin \left (\frac{\mathbf{q}\cdot (\mathbf{P}'-\mathbf{P}}{ 2} \right )] \times 
\]

\[
\left [- 
(\mathbf{S} \times \mathbf{q})\cdot \mathbf{p}(\mathbf{S} \times \mathbf{q})
-\frac{i}{ 2} (\mathbf{S}^2 \mathbf{q} - (\mathbf{S}\cdot \mathbf{q})\mathbf{S} 
)\right ] +
\]
\[
-i\frac{1 }{ 2}\frac{1 }{ (2\pi)^6}
\int d\mathbf{q} 
(\sum_i\pmb{\nabla{q}}_i [\cdot (\mathbf{S} \times \mathbf{q})_i 
v_{(\mathbf{L}\cdot \mathbf{S})^2}(\vert \mathbf{q} \vert)
e^{i \mathbf{x}\cdot (\mathbf{P}-\mathbf{P}') + i \mathbf{q}\cdot
(\mathbf{p}-\mathbf{p}')} 
(\mathbf{S} \times \mathbf{q})] \times
\]
\beq
[ (e_1-e_2) \cos \left (\frac{\mathbf{q}\cdot (\mathbf{P}'-\mathbf{P}}{ 2} \right) +i (e_1+e_2) 
\sin \left (\frac{\mathbf{q}\cdot (\mathbf{P}'-\mathbf{P}}{ 2} \right )]
\label{ap:37}
\eeq
For this interaction the term with the derivative acting on the interaction vanishes.
This is because the coefficient is
$\mathbf{q} \cdot (\mathbf{S}\times \mathbf{q})=0$.  This means that
then the  $(\mathbf{L}\cdot \mathbf{S})^2$ current matrix elements become
\[
\langle \mathbf{P}',\mathbf{p'} \vert \mathbf{J}(\mathbf{x},0)_{(\mathbf{L}\cdot \mathbf{S})^2}
\vert \mathbf{P},\mathbf{p} \rangle=
\int d\mathbf{q} 
v_{(\mathbf{L}\cdot \mathbf{S})^2}(\vert \mathbf{q} \vert)
\frac{1 }{ (2\pi)^6} e^{i \mathbf{x}\cdot (\mathbf{P}-\mathbf{P}') + i \mathbf{q}\cdot
  (\mathbf{p}-\mathbf{p}')} \times
\]
\[
[ (e_1-e_2) \cos \left (\frac{\mathbf{q}\cdot (\mathbf{P}'-\mathbf{P}}{ 2} \right) +i (e_1+e_2) 
\sin \left (\frac{\mathbf{q}\cdot (\mathbf{P}'-\mathbf{P}}{ 2} \right )]
\times
\]
\beq
\left [-  
\frac{1 }{ 2} (\mathbf{S} \times \mathbf{q})\cdot (\mathbf{p}+ \mathbf{p}')(\mathbf{S} \times \mathbf{q})
-\frac{i}{ 2}  (\mathbf{S}^2 \mathbf{q} - (\mathbf{S}\cdot \mathbf{q})\mathbf{S}) 
\right ]
\label{ap:38}
\eeq
which is equivalent to equation (\ref{ex:8}).

The part of the $\mathbf{L}^2$ current involving derivatives of the
interaction also vanishes.  To see this first note the expression for the current is
\[
\langle \mathbf{P}',\mathbf{p}' \vert 
\mathbf{J}(\mathbf{x},0)_{\mathbf{L}\cdot \mathbf{L}}
\vert \mathbf{P},\mathbf{p} \rangle =
\int d\mathbf{q} 
v_{\mathbf{L}\cdot \mathbf{L}}(\vert \mathbf{q} \vert)
\frac{1 }{ (2\pi)^6} e^{i \mathbf{x}\cdot (\mathbf{P}-\mathbf{P}') + i \mathbf{q}\cdot
(\mathbf{p}-\mathbf{p}')} \times
\]
\[
[ (e_1-e_2) \cos \left (\frac{\mathbf{q}\cdot (\mathbf{P}-\mathbf{P}'}{ 2} \right) +i (e_1+e_2) 
\sin \left (\frac{\mathbf{q}\cdot (\mathbf{P}-\mathbf{P}'}{ 2} \right )]
\]
\[
\left [
-\mathbf{q}^2\mathbf{p}
+(\mathbf{q}
\cdot \mathbf{p})
\mathbf{q}
+2  \mathbf{q}
\right ] +
\]
\[
\frac{1 }{ (2\pi)^6}
\int d\mathbf{q}
[ (e_1-e_2) \cos \left (\frac{\mathbf{q}\cdot (\mathbf{P}-\mathbf{P}'}{ 2} \right) +i (e_1+e_2) 
\sin \left (\frac{\mathbf{q}\cdot (\mathbf{P}-\mathbf{P}'}{ 2} \right )]
\times
\]
\beq
\pmb{\nabla}_q\cdot 
[
v_{\mathbf{L}\cdot \mathbf{L}}(\vert \mathbf{q} \vert)
e^{i \mathbf{x}\cdot (\mathbf{P}-\mathbf{P}') + i \mathbf{q}\cdot
(\mathbf{p}-\mathbf{p}')} \mathbf{q}^2 ]
\label{ap:39}
\eeq

\bigskip

\[
- \frac{1 }{(2\pi)^6}
\int d\mathbf{q}
[ (e_1-e_2) \cos \left (\frac{\mathbf{q}\cdot (\mathbf{P}-\mathbf{P}'}{ 2} \right) +i (e_1+e_2) 
\sin \left (\frac{\mathbf{q}\cdot (\mathbf{P}-\mathbf{P}'}{ 2} \right )]
\]
\beq
\pmb{\nabla}[\cdot( \mathbf{q}   
v_{\mathbf{L}\cdot \mathbf{L}}(\vert \mathbf{q} \vert)
e^{i \mathbf{x}\cdot (\mathbf{P}-\mathbf{P}') + i \mathbf{q}\cdot
(\mathbf{p}-\mathbf{p}')})\mathbf{q} ].
\label{ap:40}
\eeq
For a rotationally invariant
$v_{\mathbf{L}\cdot \mathbf{L}}(\vert \mathbf{q} \vert)$
\beq
(\pmb{\nabla} v) \mathbf{q}^2 = v' \hat{\mathbf{q}} \mathbf{q}^2 = 
v' q \mathbf{q}
\label{ap:41}
\eeq
while
\beq
\mathbf{q} \cdot (\pmb{\nabla}v) \mathbf{q} =
v'q\mathbf{q}.
\label{ap:42}
\eeq
These terms come with opposite signs in the expression above so
they exactly cancel.  What remains after eliminating these terms is
\[
\langle \mathbf{P}',\mathbf{p}' \vert 
\mathbf{J}(\mathbf{x},0)_{\mathbf{L}\cdot \mathbf{L}}
\vert \mathbf{P},\mathbf{p} \rangle =
\int d\mathbf{q} 
v_{\mathbf{L}\cdot \mathbf{L}}(\vert \mathbf{q} \vert)
\frac{1 }{ (2\pi)^6} e^{i \mathbf{x}\cdot (\mathbf{P}-\mathbf{P}') + i \mathbf{q}\cdot
(\mathbf{p}-\mathbf{p}')} \times
\]
\[
[ (e_1-e_2) \cos \left (\frac{\mathbf{q}\cdot (\mathbf{P}-\mathbf{P}'}{ 2} \right) +i (e_1+e_2) 
  \sin \left (\frac{\mathbf{q}\cdot (\mathbf{P}-\mathbf{P}'}{ 2} \right )]
\times
\]
\[
\left [
-\mathbf{q}^2\mathbf{p}
+  (\mathbf{q}
\cdot \mathbf{p})
\mathbf{q} -i\mathbf{q} +\frac{1 }{ 2}\mathbf{q}^2(\mathbf{p}-\mathbf{p}')
-\frac{1}{ 2} \mathbf{q}\cdot(\mathbf{p}-\mathbf{p}') \mathbf{q}  + i \mathbf{q} 
\right ] =
\]

\[
\int d\mathbf{q} 
v_{\mathbf{L}\cdot \mathbf{L}}(\vert \mathbf{q} \vert)
\frac{1 }{ (2\pi)^6} e^{i \mathbf{x}\cdot (\mathbf{P}-\mathbf{P}') + i \mathbf{q}\cdot
(\mathbf{p}-\mathbf{p}')} \times
\]
\[
[ (e_1-e_2) \cos \left (\frac{\mathbf{q}\cdot (\mathbf{P}-\mathbf{P}'}{ 2} \right) +i (e_1+e_2) 
\sin \left (\frac{\mathbf{q}\cdot (\mathbf{P}-\mathbf{P}'}{ 2} \right )]
\times
\]
%
%
%
\beq
\frac{1 }{ 2} \mathbf{q} \times (\mathbf{q} \times (\mathbf{p}+\mathbf{p}'))
\label{ap:43}
\eeq
which is equivalent to equation (\ref{ex:9}) 
Again the derivative of the potential cancels in this expression as well.
This means that the potentials can be factored.

\bibliography{master_bibfile.bib}
\end{document}